\documentclass[aps,showpacs,twocolumn,superscriptaddress]{revtex4}

\usepackage{graphicx}
\usepackage{bbm}
\usepackage{amssymb}
\usepackage{epstopdf}
\usepackage{color, amsmath, bm}
\usepackage[dvipsnames]{xcolor}
\usepackage[hidelinks]{hyperref}
\usepackage{natbib}
\RequirePackage{amsmath}
\usepackage{amsfonts}   
\usepackage{physics} 
\usepackage{comment}

\begin{document}

\title{Enhanced decoherence for a neutral particle sliding on a metallic surface in vacuum}

\author{Ludmila Viotti, Fernando C. Lombardo and Paula I. Villar}
\affiliation{Departamento de F\'\i sica {\it Juan Jos\'e
Giambiagi}, FCEyN UBA and IFIBA CONICET-UBA, Facultad de Ciencias Exactas y Naturales,
Ciudad Universitaria, Pabell\' on I, 1428 Buenos Aires, Argentina.}
\date{\today}                                           

\begin{abstract}   
Bodies in  relative motion, spatially separated in vacuum, experience a tiny friction force known as quantum friction. This force has eluded experimental detection so far due to its small magnitude and short range. Herein, we give quantitative details so as to track traces of the quantum friction by measuring coherences in the atom. We notice that the environmentally induced decoherence can be decomposed into contributions of different signature: corrections induced by the electromagnetic vacuum in presence of the dielectric sheet and those induced by the motion of the particle. In this direction, we show that non-contact friction enhances the decoherence of the moving atom. Further, its effect can be enlarged by a thorough selection of the two-level particle and the Drude-Lorentz parameters of the material.  In this context, we suggest that measuring decoherence times through velocity dependence of coherences could indirectly demonstrate the existence of quantum friction.
\end{abstract}

\maketitle

\section{Introduction}
Some outstanding features of modern quantum field theory are the nontrivial structure exhibited by the \mbox{vacuum} state and  consequent existence of \mbox{vacuum} \mbox{fluctuations}. These quantum fluctuations induce \mbox{macroscopic} effects over which, in many cases, \mbox{experimental} verification has been achieved and thereafter improved. The most renewed example is the Casimir static force between neutral bodies placed in \mbox{vacuum} \cite{Casimir, book_milonni, Lamoreaux1997, bordag1, bordag2, book_milton, milton2004casimir, reynaud2001quantum}. Another paradigmatic example is that of the phenomenon known as dynamical Casimir effect (DCE) \cite{review_friction,nation_colloquium,dyncasexp_supercond,dyncasexp_squid} in which a \mbox{mirror} moving at time dependent velocities produces real photons through quantum vacuum fluctuations.
However, while experimental observation has been attained on these effects, another phenomenon has eluded experimental detection so far due to its short range and small magnitude. This latter effect consists on the appearance of a dissipative force between spatially separated objects in relative motion, known as quantum friction (QF) \cite{pendry97,pendry_debate,Pendry_reply, Leonhardt_2010, Leonhardt_nofriction, volokitin_persson}. This lack of experimental verification favors the co-existence of different \mbox{theoretical} approaches which rely on a variety of assumptions and do not converge to a single result \cite{dalvit_arbitrarily}. Such variety of methods ranges from time-dependent perturbation theory \cite{barton_atom_halfspace, intravaia_acceleration}, quantum master equations in markovian limit \cite{Scheel} to generalized non-equilibrium fluctuation-dissipation relations \cite{dalvit_fluctuation} and thermodynamic principles \cite{hu_thermodynamic}. There have also been a great theoretical effort devoted into finding favorable conditions for experimental measurements of QF \cite{farias2018quantum, carusotto2017friction,intravaia_rolling,farias_graphene,farias_friction,viotti_thermal}. 

Due to the experimental challenges involved on the implementation of precision measurements for the detection of such a small force on objects near a surface, lately there has been a set of works following a different approach consisting on tracking traces of quantum friction through the study of velocity-dependent quantum-vacuum effects which could more easily testable.  In Refs. \cite{volokitin2011quantum, volokitin_cherenkov}, authors have 
investigated the Van der Waals friction between graphene and an amorphous SiO2 substrate. They found the electric current to saturate at a high electric field due to this friction. The saturation current depends weakly on the temperature, which they attributed to the quantum friction between the graphene carriers and the substrate optical phonons. They calculated the frictional drag between two graphene sheets caused by Van der Waals friction, and proved that this drag can induce a voltage high enough to be measured experimentally by state-of-art non-contact force microscopy. This work paved the way for a possible mechanical detection of the Casimir friction. In  \cite{buhmann_spectroscopic}, the level shift and decay rate modification arising from the motion of an atom in presence of a medium are found in markovian limit, and their relation to the parallel component of vacuum force is discussed. Considering other aspects of the internal dynamics, a toy model for  decoherence induced on the state of a particle in relative motion to a material (\mbox{modeled} as set of harmonic oscillators) has been studied in \cite{farias_decoherence}. Recently, in \cite{farias_nature} some of us have examined the effect of the vacuum, dressed by the presence of a more realistic Drude-Lorentz material on the geometric phase acquired by an atom traversing at constant velocity  and made a proposal for an experimental setup. 

Following these ideas, in this article 
we shall focus on finding an alternative indicator of the existence of 
quantum friction and explore favorable experimental conditions under which the quantum friction can be detected. Due to the technical issues explained above, we shall \mbox{follow} a different approach consisting on  tracking traces of quantum friction in the coherences of a two-level system.
We will pursue this goal by a thorough study of the  decoherence process suffered by
 a neutral particle in non relativistic motion parallel to a  metallic surface in electromagnetic vacuum. Particularly, we shall try to identify  environmental induced behaviour dependent exclusively on the particle's velocity since the mere presence of a velocity contribution in the noise induced corrections is a testimony of the existence of frictional effect. Environmental induced decoherence can be decomposed in two contributions: corrections induced by the solely electromagnetic vacuum in presence of the dielectric sheet and those induced by the motion of the particle. In the end, we aim to prove that the presence of velocity and hence, non-contact friction enhances the decoherence of the internal degrees of freedom of the moving atom, suggesting that measuring decoherence times could be used to indirectly demonstrate the existence of quantum friction. 
The article is organized as follows: in Sec. \ref{system_section} we provide a description of the composite system under investigation.  In Sec. \ref{evolution_seccion}, we solve the complete system's dynamics in the non retarded regime and weak coupling limit without making either markovian or low dissipation approximations. In Sec. \ref{destruction_seccion} we present a complete analysis of the environmental induced dynamics of the system so as to track evidence of quantum fluctuation induced effects due to the velocity of the particle, focusing on the conditions under which those effects are enhanced. This research is mainly conducted by observing the suppression of the coherences of the internal degree of freedom of the particle, where this destruction is found to be fastened by the movement of the particle. We further include an analysis of different materials and particles and study how these features impact on the magnitude of the effects under study. In Sec. \ref{conclu} we summarize our main conclusions. Two appendices complement the work.


\section{The system}\label{system_section}

\begin{figure}[t]
\centering
\includegraphics[width=.75\columnwidth]{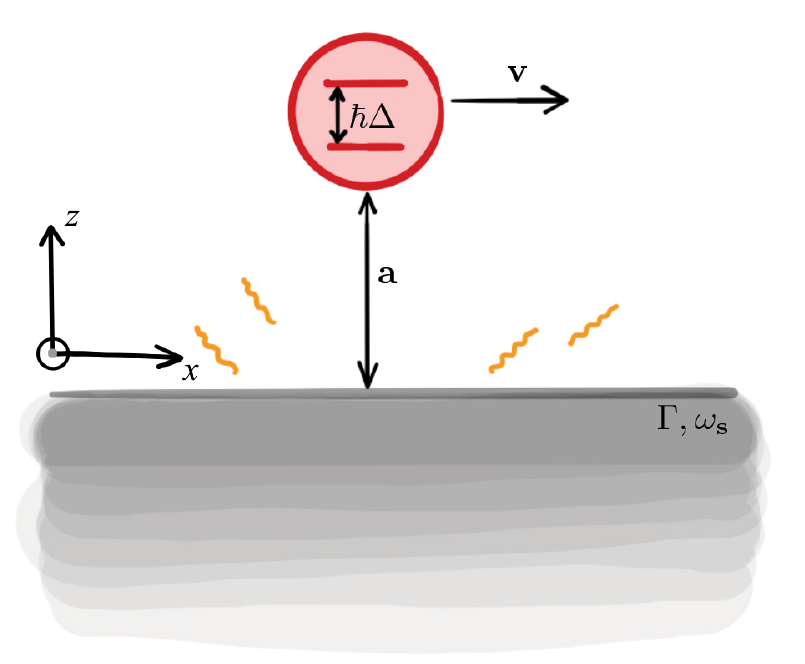}
\caption{\label{esquema} A scheme of the system under consideration, where the two-level system moves at a fixed distance $a$ from dielectric plate.}
\end{figure}

Here we consider a neutral particle moving through medium-assisted electromagnetic field vacuum. As shown in Fig.\ref{esquema}, the particle is modeled as a two level system whose center of mass follows a prescribed trajectory  $\mathbf{r}_{\text{s}}(t) = v\,t\;\check{x} + a\;\check{z}$ at a fixed distance $a$ from a dielectric semi infinite planar medium. At this point it is worth noting that we have employed the inverted hat to denote unit vectors as to save the regular hat to denote operator nature.  The dynamics of the composite system can be described by a Hamiltonian consisting on atomic, field and interaction contributions defined by:
\begin{equation}
    \hat{H}=\frac{\hbar}{2}\Delta\; \hat\sigma_\text{z} \otimes \mathbbm{1} + \hat{H}_{\text{em}}+\hat{H}_{\text{int}},
\end{equation}
where $\Delta$ is energy gap of the two level system and $\hat{H}_{\text{em}}$ is the Hamiltonian of the electromagnetic field in absence of the particle, but  in presence of the dielectric half space at $z<0$. The interaction between the particle and the field is given in dipole (long wavelength) approximation by $\hat{H}_{\text{int}} = - \hat{\mathbf{d}}\,\otimes\, \hat{\mathbf{E}}(\mathbf{r}_{\text{s}})$,
and depends  explicitly  on time through the position of the particle, which is treated as a classical variable relying on its uncertainty to be unresolvable by the characteristic wavelength of the electric field.
We shall restrict ourselves to the non-retarded (near field) regime where the particle-surface distance $a$ is small enough to satisfy $a \Delta/c\ll1$. In this regime, the finite time taken for a reflected photon to reach the particle is negligible compared to its natural timescale and the interaction Hamiltonian can therefore be written as $\hat{H}_{\text{int}} = \hat{\mathbf{d}}\otimes \nabla\hat{\Phi}(\mathbf{r}_{\text{s}})$, where the electric potential $\hat{\Phi}$, expanded in plane-wave basis corresponding to elementary excitations is \cite{barton78,barton79}
\begin{equation}
    \hat{\Phi}=\int d^2k\;\int_0^\infty d\omega \left(\hat{a}_{\mathbf{k},\omega},\phi(\mathbf{k},\omega)e^{i \mathbf{k} \mathbf{r}_\parallel} + h.c.\right),
\end{equation}
and contains all the information of the electric field in the $z>0$ region, dressed by the dielectric medium. The bosonic operators satisfy the commutation relation  $\left[\hat{a}_{\mathbf{k},\omega},\hat{a}^\dagger_{\mathbf{k}',\omega'}\right]=\delta\left(\mathbf{k}-\mathbf{k}'\right)\delta\left(\omega-\omega'\right)$, and the single excitation mode functions are given by

\begin{equation}
    \phi\left(\mathbf{k}, \omega\right)=\sqrt{\frac{\omega\Gamma}{\omega_\text{s}}}\sqrt{\frac{\hbar}{2\pi^2 k}}e^{-k\,z}\frac{\omega_{\text{p}}}{\omega^2-\omega_{\text{s}}^2 - i \omega \Gamma},
\end{equation}
where the wave vector $\mathbf{k}=(k_x, k_y)$ is parallel to the medium surface and $k=|\mathbf{k}|$. The frequency $\omega_{\text{s}}$ gives the surface plasmon resonance and the material dissipation rate $\Gamma$ its broadening while, in Drude model for metals, the plasma frequency $\omega_{\text{p}}$ satisfies $\omega_{\text{p}}^2= 2\omega_{\text{s}}^2$.


\section{Non-unitary evolution of the system}\label{evolution_seccion}
In order to address the dynamics of the two level system, we  shall derive the master equation satisfied by the reduce density matrix representing its state. This is done  by integrating out the degrees of freedom corresponding to the composite environment, as indicated by  the formalism of open quantum systems \cite{petruccione}. By assuming an initially factorized state $\rho(0) = \rho_{\text{s}}(0)\,\otimes\,\rho_{\text{em}}^{\text{vac}}$ with the dressed electromagnetic field in its vacuum state, the master equation in the interaction picture, up to second order in the coupling constant is given by \cite{paz_mastereq}

\begin{equation}
    \dot{\rho}_\text{s}(t)=\frac{-1}{\hbar^2}\int_0^t dt'\, \Tr_{\text{em}} \left[V(t),\left[V(t'),\rho_{\text{s}}(t)\otimes \rho_{\text{em}}\right]\right].
    \label{meq_formal}
\end{equation}
An explicit computation of the above expression leads to the equation ruling the temporal evolution of the reduced density matrix \cite{Leggett,book_referee2}.
In this work we have considered the equation governing the two-level system dynamics that results from performing the secular approximation, also referred as post-trace rotating wave approximation. This approximation consists on neglecting those terms which are fast-oscillating in the interaction picture and it can be performed based on the assumption that dissipative corrections are weak enough expecting to preserve accurate results in the timescales of the phenomena \cite{hu_rwa,maniscalco3}. Therefore, the equation we obtain is

\begin{align}\nonumber
    \dot{\rho}_{\text{s}}=&-\frac{\mathfrak{i}\Delta}{2}\left[\hat{\sigma}_z,\rho_{\text{s}}\right]+\mathfrak{i}\,\zeta(v,t)\left[\sigma_x,\{\sigma_y,\rho_{\text{s}}\}\right]\\[.75em]\nonumber
    &-\frac{1}{2} D(v,t)\left(\left[\sigma_x,[\sigma_x,\rho_{\text{s}}]\right]+\left[\sigma_y,[\sigma_y,\rho_{\text{s}}]\right]\right)\\[.75em]
    &-\frac{1}{2}f(v,t)\left(\left[\sigma_x,[\sigma_y,\rho_{\text{s}}]\right]-\left[\sigma_y,[\sigma_x,\rho_{\text{s}}]\right]\right)
    \label{meq_secular}
\end{align}
where the non-unitary effects are modeled by the diffusion coefficients $D(v, t)$ and $f(v, t)$, while dissipative effects are present in $\zeta(v, t)$. All three coefficients consist on real functions of time, with parameters introduced by the particle and the medium assisted field. These coefficients are developed in Appendix \ref{appendixa}, where an analytical solution is given for sufficiently small velocities of the particle.

It is important to mention that the diagonal elements of the reduced density matrix are exactly the same whether secular approximation is performed or not, since for this system it only implies disregarding a dynamical interaction between $\rho_{12}$ and $\rho_{21}$. By resorting to a change of variables, say $\rho_-=\rho_{11}-\rho_{22}$ and $\rho_+=\Tr(\rho_{\text{s}})=1$,  a formal solution can easily be found through direct computation,
\begin{align}\nonumber
    \rho_-(t)=&e^{-4\int dt \;D(v.t)}\rho_{-}(0)\\
    &-4e^{-4\int dt \;D(v.t)}\; \int_0^t dt'\;\zeta(v,t')e^{4\int dt' \;D(v.t')}.
    \label{rmenos}
\end{align}
The non-diagonal elements in this approximation are,
\begin{align}
    \rho_{12}^{\;\text{SEC}}(t)&=\rho_{12}(0)\;e^{-\int_0^t\;dt'\left(2D(v,t')+2\mathfrak{i}f(v,t') + \mathfrak{i}\Delta\right)}\\[1em]
    \rho_{21}(t)&=\rho_{12}^*(t).
    \label{r12_sec_expresion}
\end{align}
Hence, after applying the secular approximation, the reduced density matrix describing the state of the particle's internal degree of freedom can be written as $ \rho_{\text{s}}(t)=\rho_{\text{diag}} + \rho_{\text{nond}}$, with 
\begin{equation}
  \rho_{\text{diag}}= \left( \begin{array}{cc}
      \rho_{11}(t)   &  0\\
       0  & 1-\rho_{11}(t)
    \end{array}\right),
    \label{rho_diag}
\end{equation}
and
\begin{equation}
   \rho_{\text{nond}}= \left(\begin{array}{cc}
      0   & \rho_{12}(0)e^{-\mathfrak{i}\xi(t)}\\
      \rho_{12}^*(0)e^{\mathfrak{i}\xi(t)} & 0
    \end{array}\right)e^{-\frac{ 2}{\omega_{\text{s}}}\int_0^t dt'\;D(v,t')},
    \label{rho_nond}
\end{equation}
with 

\begin{equation}
    \xi(t)=\mathfrak{i} ,\tilde{\Delta}\,t +2\mathfrak{i}\int_0^t dt'\;\frac{f(v,t')}{\omega_{\text{s}}},
\end{equation}
where we have used the dimensionless parameters $t= \omega_{\text{s}}\,t_{\text{real}}$, $u= v/(\omega_{\text{s}}\,a)$, $\tilde{\Gamma}= \Gamma/\omega_{\text{s}}$ and $\tilde{\Delta}= \Delta/\omega_{\text{s}}$ as defined in Appendix \ref{appendixa}. 

\begin{figure}[h]
\centering
\includegraphics[width=.95\columnwidth]{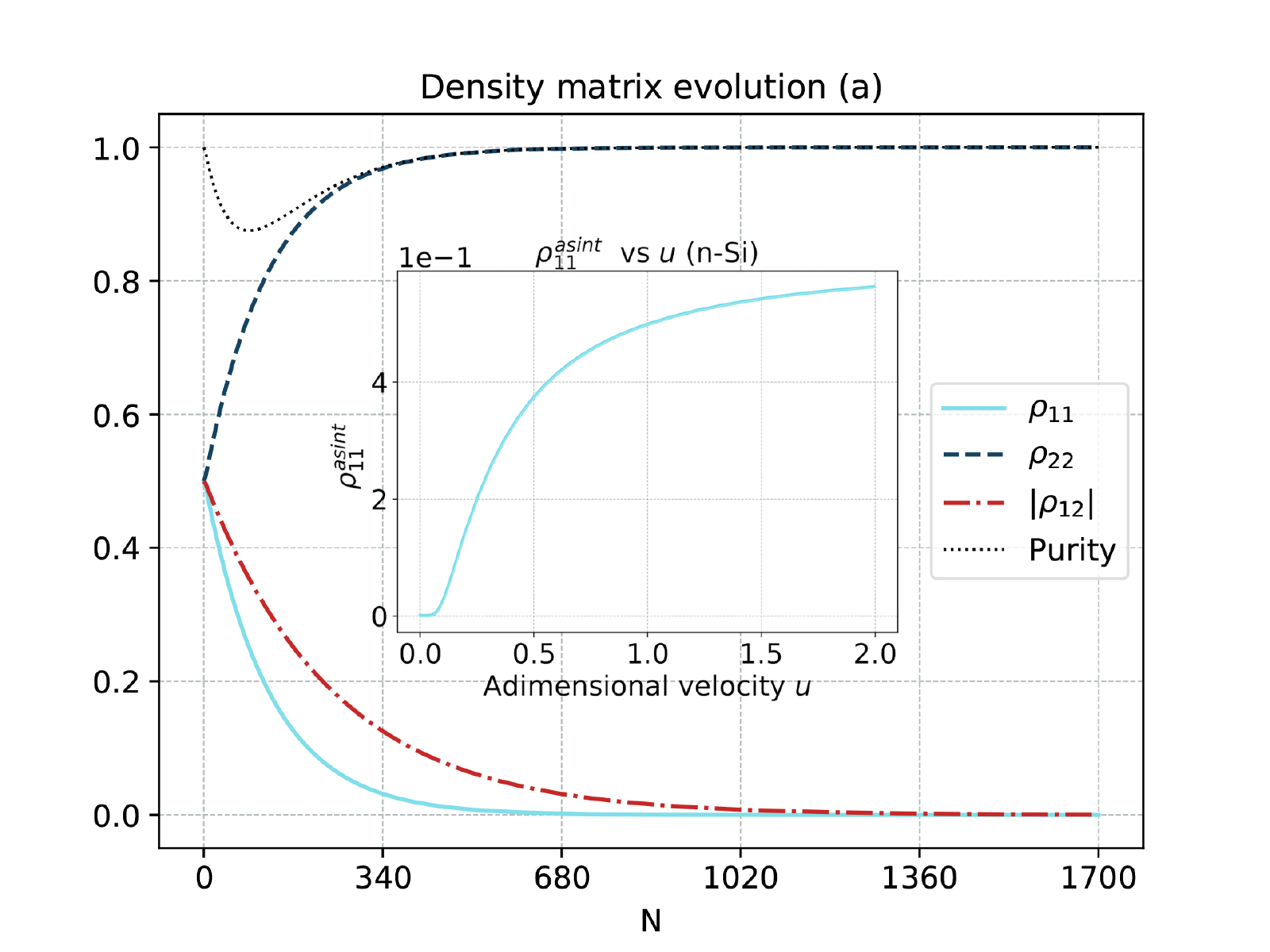}

\vspace{-.2cm}
\includegraphics[width=.95\columnwidth]{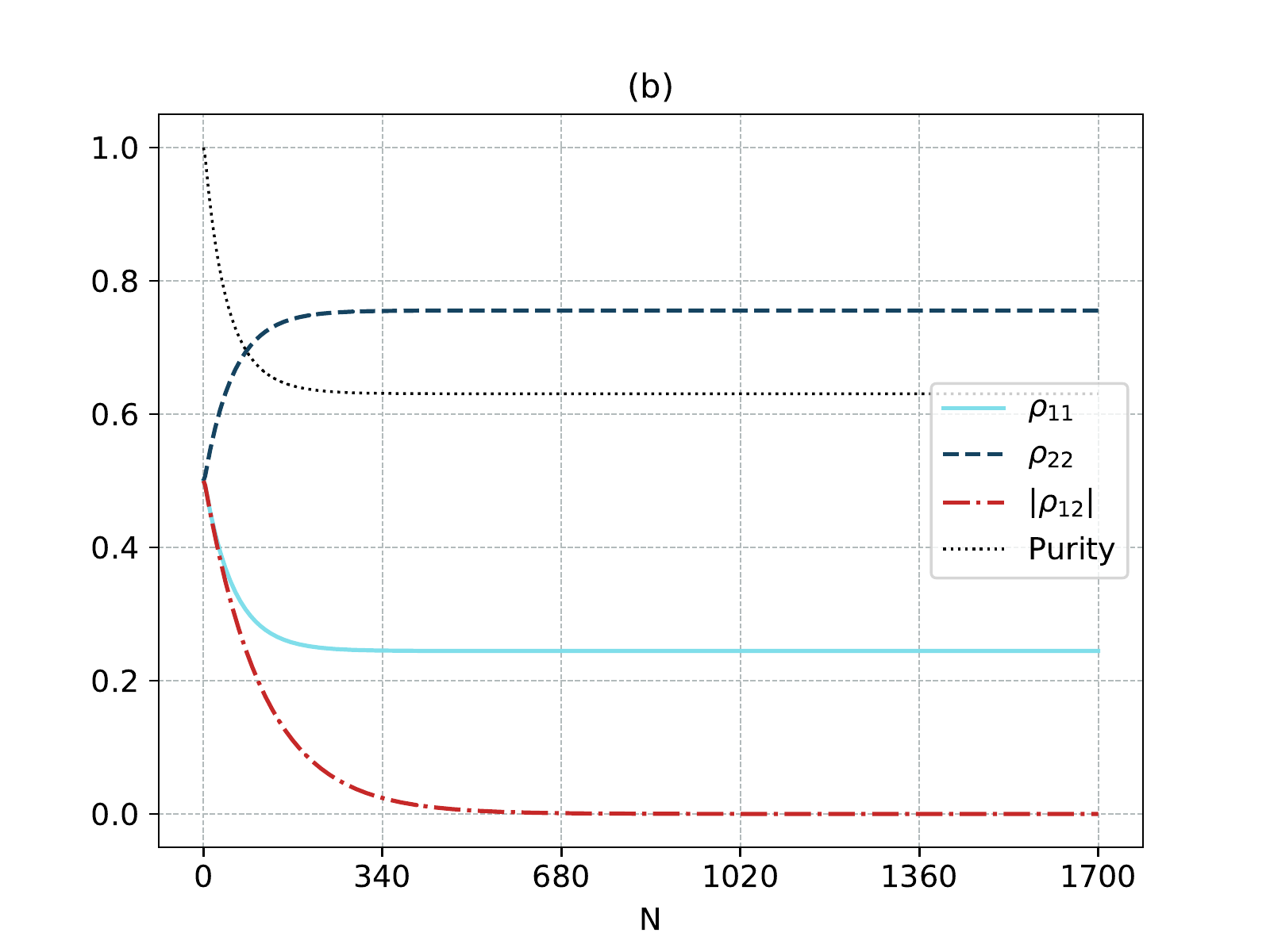}
\caption{Matrix elements and purity evolution, in natural cycles $N = \frac{t}{2\pi/\tilde{\Delta}}$ . (a) The system can be seen to tend to its ground state independently of the initial state for small velocities. (b) For large enough velocity, the asymptotic state is a mixed state. Parameters values are $\tilde{\Gamma}=1$, $r_0/\omega_\text{s}= 10^{-2}$, $\tilde{\Delta}=0.2$, $u= 0.003(a)/0.3(b).  $ \label{matriz_tiempo}. The insert in (a) displays the behaviour of the asymptotic value of $\rho_{11}$ with dimensionless velocity $u$. }
\end{figure}

The dynamics of the system can be seen to display two qualitative different behaviors depending on the velocity of the particle. While for large enough velocities the evolution leads to a mixed asymptotic state, if the velocity is small the system evolves to its ground state. We present both behaviors in Fig. \ref{matriz_tiempo}. In Fig. \ref{matriz_tiempo}(a)  the particle is considered to move with dimensionless velocity $u=0.003$. In that case, both the coherences and the $\rho_{11}$ element of the state are indefinitely suppressed and tend to vanish for large enough times leading the system to its ground state. The purity of the state decreases up to reaching a minimal value from which it starts to recover and finally tends to the unity as system tends to its pure ground state. On the other hand, while the coherences are totally extinguished, for a dimensionless velocity $u=0.3$ the $\rho_{11}$ element of the reduced density matrix is only suppressed up to an asymptotic value. This behavior, collected in Fig.\ref{matriz_tiempo}(b), leads to a mixed asymptotic state whose purity never recovers but lands to a constant value depending on $\rho_{11}$ as $p=\rho_{11}^2+(1-\rho_{11})^2$.

\begin{figure}[h]
\centering
\includegraphics[width=.95\columnwidth]{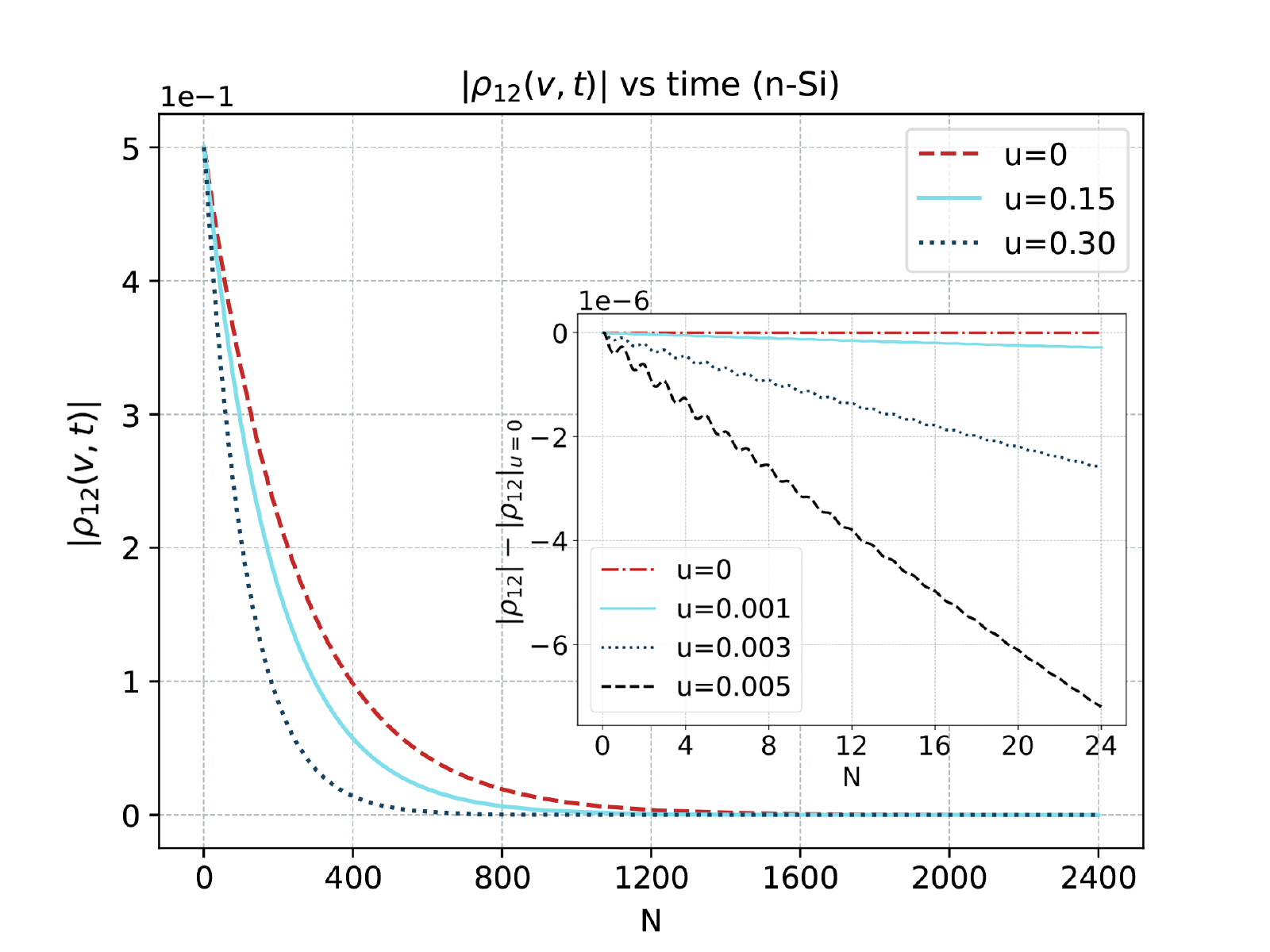}
\caption{Coherence evolution, in natural cycles $N = \frac{t}{2\pi/\tilde{\Delta}}$, for different velocity values. The insert shows the difference in the coherence as cycles goes by for smaller velocity values. Parameters values are $\tilde{\Gamma}=1$, $r_0/\omega_\text{s}=10^{-2}$, $\tilde{\Delta}=0.2$.\label{r12_tiempo}}
\end{figure}

The insert in Fig. \ref{matriz_tiempo}(a) shows the velocity dependence of asymptotic value of $\rho_{11}$. The dimensionless velocity $u = \tilde{\Delta}/2$ at which the dynamics of the system acquires a completely different behaviour coincides with lower bound on the velocity which allows the atom to become excited at the expense of its kinetic energy  \cite{dalvit_nonmarkovianity, plasmones, cherenkov}. A few remarks can be mentioned from a more detailed observation of the coherences behaviors. Fig. \ref{r12_tiempo} shows off-diagonal elements of the density matrix to be suppressed by the environment. This destruction is not only seen to be fastened by the relative motion between the particle and the material, but also to happen sooner as the velocity is increased, as for example, the absolute value of the coherence for a particle in relative motion with $u=0.3$ extincts sooner than that with $u=0.15$. The same monotone behavior is displayed in the insert, in which the difference between absolute values $|\rho_{12}(t)|-|\rho_{12}^{u=0}(t)|$ grows faster as the velocity is enlarged.

However, as the difference between the value of $|\rho_{12}(t)|$ for finite velocity and that for null velocity gets a maximum value and tends to vanish afterwards, explorations of the coherences at too late times would not allow any identification of the velocity induced effects. As shown in Fig. \ref{matriz_tiempo}, the same observation would apply to the behavior of both the populations and purity measurements for small velocities as all these converge to a given value for long enough times, independently of the velocity. For large velocities, however, the velocity effect could be observed at late times on the populations or purity measurements, while the coherences will always be completely destructed if its waited too long.  
In the following, we shall use the results herein obtained for the reduced density matrix so as to define a scale in which coherences in the internal degree of freedom of the atom, are destroyed by the influence of the electromagnetic field dressed by the metallic material.


\section{Environment induced destruction of coherences}\label{destruction_seccion}

In this section we shall focus on the environment induced destruction of the particle's quantum coherences and define a characteristic time in which the process takes place. 
The density matrix defined by Eqs.(\ref{rho_diag}) and (\ref{rho_nond}) allows to define a decoherence timescale $\tau_{\text{D}}$ from the decoherence function ${\cal D}(t) = e^{-\frac{ 2}{\omega_{\text{s}}}\int_0^t dt'\;D(v,t')}$ as ${\cal D}(\tau_{\text{D}}) = e^{-2}$.


For small velocities, up to second order in the dimensionless velocity $u$ this decoherence timescale behaves as
\begin{align}\nonumber
    \tau_{\rm D}= \tau_{\rm D}^\text{Mark} + \left[\frac{-1}{\sqrt{4-\tilde{\Gamma}}^2} \frac{g(\tilde{\Delta},\tilde{\Gamma})}{h(\tilde{\Delta},\tilde{\Gamma})}+ \frac{2}{\pi \tilde{\Delta}}\right]\\[0.75em]\nonumber
    + \frac{3}{8}\frac{d^{(a)}}{d^{(i)}}u^2 \times \left\lbrace\left[ g(\tilde{\Delta},\tilde{\Gamma}) \frac{\partial_{\tilde{\Delta}}^2h(\tilde{\Delta},\tilde{\Gamma})}{h^2(\tilde{\Delta},\tilde{\Gamma})}-\frac{\partial_{\tilde{\Delta}}^2g(\tilde{\Delta},\tilde{\Gamma})}{h(\tilde{\Delta},\tilde{\Gamma})} \right]\right.\\[0.75em]
    \left. + \frac{2}{\pi h(\tilde{\Delta},\tilde{\Gamma})}\left[\partial_{\tilde{\Delta}}^2\frac{h(\tilde{\Delta},\tilde{\Gamma})}{\tilde{\Delta}}-\frac{\partial_{\tilde{\Delta}}^2h(\tilde{\Delta},\tilde{\Gamma})}{\tilde{\Delta}}\right]\right\rbrace,
    \label{tdec_tlargo}
\end{align}
where the term corresponding to the markovian approximation can be expressed as

\begin{equation}
    \tau_{\rm D}^\text{Mark} = \frac{\hbar \omega_{\text{s}}^2\,a^3}{d^2\omega_{\text{p}}^2}\frac{32}{d^{(i)}}\left(\frac{1}{h(\tilde{\Delta},\tilde{\Gamma})}-\frac{3}{8}\frac{d^{(a)}}{d^{(i)}}u^2\frac{\partial_{\tilde{\Delta}}^2h(\tilde{\Delta},\tilde{\Gamma})}{h^2(\tilde{\Delta},\tilde{\Gamma})}\right)
    \label{tdec_mark}.
\end{equation}

The functions $h(\tilde{\Delta},\tilde{\Gamma})$ and $g(\tilde{\Delta},\tilde{\Gamma})$ appearing in these expressions are defined by 

\begin{align}\nonumber
    h(\tilde{\Delta},\tilde{\Gamma}) &= \frac{\tilde{\Delta}\tilde{\Gamma}}{(\tilde{\Delta}^2-1)^2+\tilde{\Delta}\tilde{\Gamma}}\\[.75em]\nonumber
    g(\tilde{\Delta},\tilde{\Gamma}) &=\Re\left[\left(1+\frac{2\mathfrak{i}}{\pi}\log(\tilde{\omega}_{\text{r}}/\tilde{\Delta})\right)\right.\\
    &\hspace{1cm}\left.\left(\frac{1}{(\tilde{\omega}_{\text{r}}+\tilde{\Delta})^2}+\frac{1}{(\tilde{\omega}_{\text{r}}-\tilde{\Delta})^2}\right)\right]
\end{align}
with $\tilde{\omega}_{\text{r}}$ as defined in Eq.(\ref{raiz}) and the dependence on the polarization orientation encoded in $d^{(i)}$ and $d^{(a)}$ as defined after Eq. (\ref{orientacion}). The dependence of the dynamics on the velocity of the particle, (which is already evident in the behavior displayed in Figs. \ref{matriz_tiempo} and \ref{r12_tiempo}), can therefore be studied by the use of the decoherence time $\tau_{\text{\rm D}}$, which happens to scale as $u^2$ for low velocities as seen from Eqs. (\ref{tdec_tlargo}) and  (\ref{tdec_mark}). This quadratic behavior of the internal dynamics is in agreement with the results found in \cite{buhmann_spectroscopic}, where among other aspects of the internal dynamics of an atom, the decay rate - which is proportional to the markovian limit of $D(v,t)$ - is found to scale as $u^2$. A similar dependence with $u$ was found in \cite{farias_nature} for the corrections induced by the non-contact friction force on the accumulated geometric phase. 

\begin{figure}[h]
\centering
\includegraphics[width=\columnwidth, trim = 0 20 0 20 ]{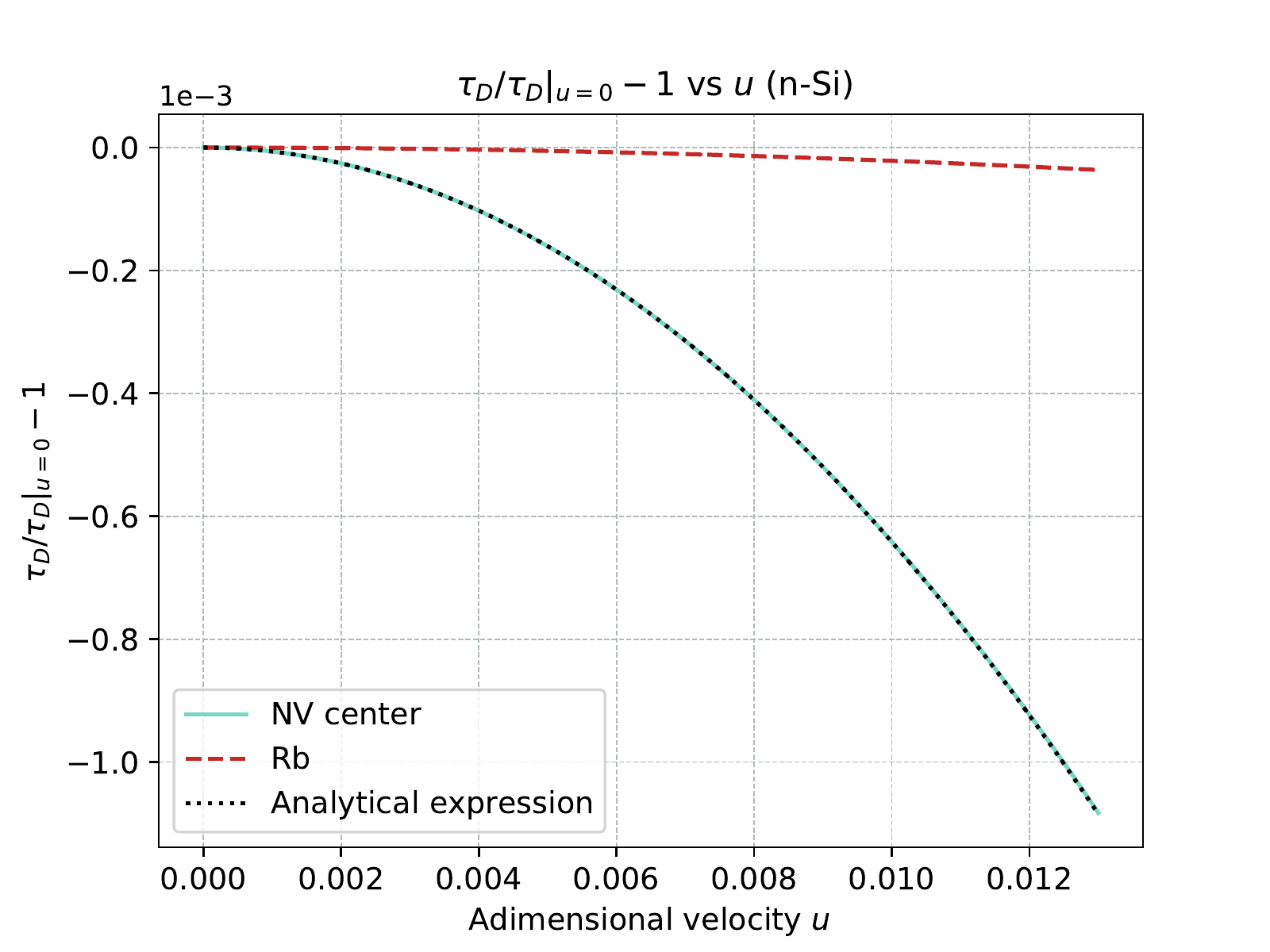}
\vspace{-.2cm}
\caption{Decoherence time rate as a function of the adimensional velocity $u$. Parameters values are $\tilde{\Gamma}=1$,$\tilde{\Delta}^{NV}=0.2$/$\tilde{\Delta}^{Rb}=8$, $r_0/\omega_\text{s}=10^{-2}$. \label{tdec_vs_v}}
\vspace{-.5cm}
\end{figure}

Decoherence time being approximated by $\tau_{\rm D}\sim a - b\,u^2$ reveals the effect of the environment on the particle contains two  contributions of different nature: (i) a contribution induced by solely vacuum fluctuations (dressed by the presence of the dielectric) and (ii) a contribution induced by the motion of the particle in quantum vacuum. Then, it results instructive to study the factor $b/a$ as it constitutes
a rate among these two contributions. We are defining $\tau_{\rm D}$ as the net effect of the environment on the particle while $\tau_{\rm D}|_{u=0}$ is the decoherence time when the particle is static. If velocity effects are insignificant $\tau_{\rm D}/\tau_{\rm D}|_{u=0} \sim 1$. Hence, by inspecting the rate 
$(\tau_{\rm D}/\tau_{\rm D}|_{u=0} -1)$ we gain access to $b/a$.
In Fig. \ref{tdec_vs_v}, this rate is plotted as a function of velocity for two different $\tilde{\Delta}$ values where a  quadratic behavior can be easily noted. This behavior  is also confirmed  when  compared  the rate computed of Eq. (\ref{tdec_tlargo}) and the one numerically obtained from the proper definition of $\tau_{\rm D}$.

The difference in the scale factor on each curve reflects how both (the net effect of the environment on the particle internal degree of freedom and the contribution derived from the finite velocity) are strongly dependent on the parameters of the problem, which are introduced by the material of the half-space, the level-spacing of the particle and its velocity. For example, the relation $b/a$ takes a numerical value $b/a \sim 6.417$ for an NV-center moving over a n-doped silicon (n-Si) surface, while it takes a value $b/a \sim 0.216$ for a rubidium (Rb) atom moving over the same surface.

\begin{figure}[h]
\centering
\includegraphics[width=.9\columnwidth, trim= 20 20 20 20]{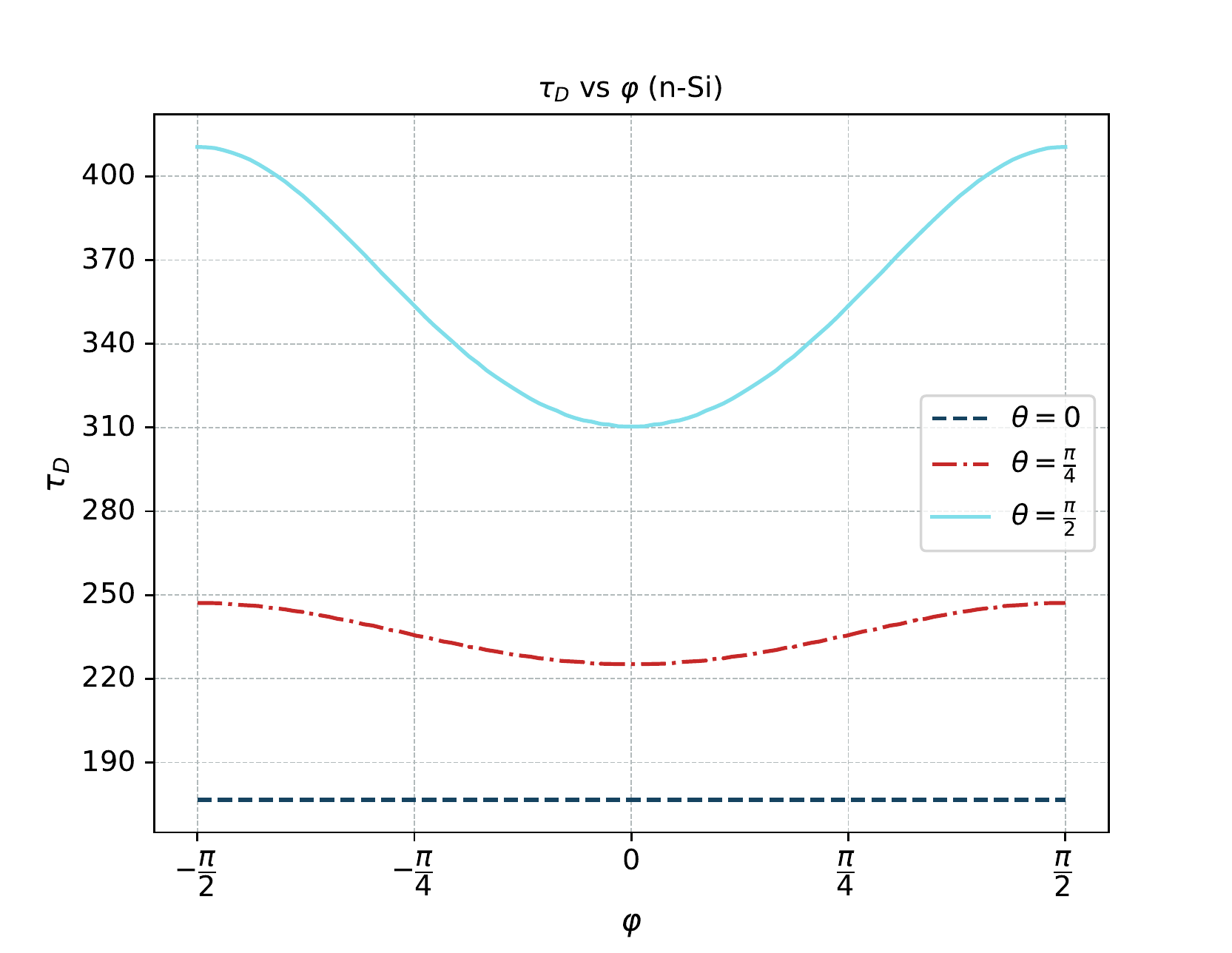}
\caption{Decoherence time as a function of the polarization direction of the system. Parameters values are $\tilde{\Gamma}=1$, $\tilde{\Delta}=0.2$, $u=0.3$ and $r_0/\omega_\text{s}=10^{-2}$. \label{tdec_vs_phi_fantasia}}
\vspace{-.1cm}
\end{figure}

The timescale defined in this way inherits also a dependence with the orientation of the polarization of the system $\mathbf{d}=d(\sin(\theta)\cos(\varphi)\hat{x} + \sin(\theta)\sin(\varphi)\hat{y} + \cos(\theta)\hat{z})$ (where $\varphi$ and $\theta$  are the spherical azimuthal and polar angles, respectively) from the coefficients governing the dynamics. Fig.\ref{tdec_vs_phi_fantasia}  shows the $\varphi$ dependence for different fixed $\theta$ values, where it can be seen that the decoherence time gets its smallest value when the polarization is perpendicular to the dielectric surface. If tilted, the coherences fall sooner when the polarization is in the direction of the velocity. This behavior is in accordance to that shown by Intravaia et al. in the insert of Fig.5  in \cite{dalvit_nonmarkovianity} where they have shown the frictional force (computed up to second order in the coupling constant) dependence with the polarization direction. 
Herein,  we find that for the same dipole orientation the force increases, $\tau_{\text{D}}$ decreases, implying that  decoherence effects are stronger in that case.  This permits a direct link between decoherence and quantum friction since they exhibit a qualitative inverse proportionality: the more decoherence effect,  the bigger the frictional force.

\begin{figure}[h]
\centering
\includegraphics[width=.9\columnwidth, trim= 20 20 20 10]{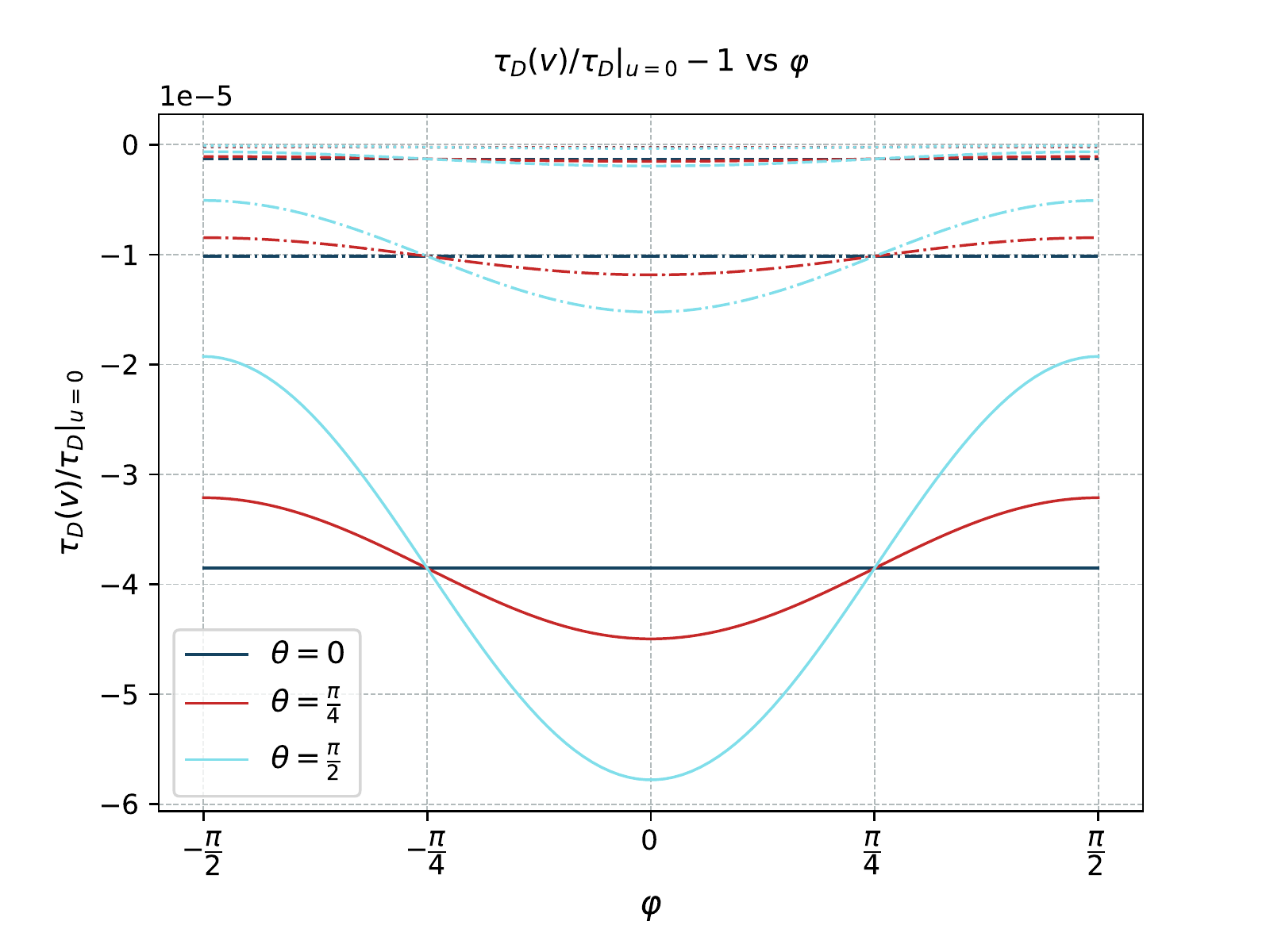}
\caption{Decoherence time as a function of the polarization direction of the system, which has polarization $\mathbf{d}=d(\sin(\theta)\cos(\phi)\hat{x} + \sin(\theta)\sin(\phi)\hat{y} + \cos(\theta)\hat{z})$ both for different systems and different materials. Different linestyles represent different system-material combinations in this correspondence: Dotted lines correspond to Rb atom-Au, dashed lines correspond to Rb atom - n-Si, dot-dashed lines to NV center-Au and solid lines correspond to NV center - n-Si. \label{tdec_vs_phi}}
\end{figure}

Seeing in Fig.\ref{tdec_vs_v} that those variations introduced by the different parameters concerning the material and particle seem to be relevant for the magnitude of the effect, we complete this section examining some possibilities. For the dielectric material we shall consider a dielectric to be a gold surface (Au) or  an n-doped silicon material (n-Si), gold having parameters of Drude-Lorentz model $\omega_{\text{s}}^{\text{Au}}\sim 9.7 \times 10^{15}$ rad/s and $\Gamma/\omega_{\text{s}}\sim 0.003$  while n-Si parameters are $\Gamma/\omega_{\text{s}}\sim 1$ and $\omega_{\text{s}}^{\text{n-Si}}\sim 2.47 \times 10^{14}$ rad/s. As for the particles (atoms), we shall consider a Rb atom or a single NV center in diamond as an effective two-level system. In Fig.\ref{tdec_vs_phi}, we show the behavior of $\tau_{\rm D}/\tau_{\rm D}|_{u=0}-1$ on the polarization direction for different sets of frequencies. Therein, we include all four combinations: dotted lines represent the decoherence time  Rb atom - Au, dashed lines correspond to Rb atom- n-Si,  dot-dashed lines to NV center- Au and solid lines correspond to NV center - n-Si. In this way, we can get an insight of the importance of the velocity dependent effects since the bigger the magnitude of the quantity displayed, the more important the $u^2$ contribution becomes.  The results obtained enhance the idea that the velocity effects induced on the atom depend on the material and particle considerably. The rate between the decoherence timescale at finite velocity and that at null velocity is enlarged by a factor $10^2$ when comparing an NV center moving over a n-Si coated surface with a Rb atom moving over a gold coated surface.

We conclude from our analysis that the election of n-Si as the material and an NV center as our system would enhance the effect the most. The NV center consists of a vacancy, or missing carbon atom, in the diamond lattice lying next to a nitrogen atom, which has substituted for one of the carbon atoms. The electron spin is the canonical quantum system and the NV center offers a system in which a single spin can be initialized, coherently controlled, and measured. It is also possible to mechanically move the NV center. For a deeper discussion on the dependence of the effect with the natural level spacing of the system we refer to Appendix (\ref{appendixb}). 

\section{Conclusions}\label{conclu}

In this article, we have studied the complete dynamics of a two-level system in relative motion to a semi-infinite dielectric material in the electromagnetic vacuum field, with the purpose of characterizing the effects of motion in the dynamics as an alternative to the explicit computation of QF. We have derived the perturbative master equation and obtained the reduced density matrix for all times, we have done this in the near-field regime but otherwise without referring to Markov approximation. Further, we have obtained an analytical expression in the low velocity limit when computing the environmental kernels for the vacuum fluctuations. 

By a direct inspection of the density matrix evolution, completely different behavior has been found depending on the particle's relative velocity to the dielectric surface. While for low velocities both $\rho_{11}$ and $\rho_{12}$ elements are completely suppressed and the system finally tends to its ground state, when the velocity increases the environment effect produces mixed asymptotic states, as the element $\rho_{11}$ does not completely vanish but lands to a finite constant value. This divergence on the asymptotic states leads to a completely different behavior of the state purity, which is initially reduced but afterwards recovers to unity when the asymptotic state is pure ground state but is permanently reduced to a lower value when the velocity is large enough as to allow for mix asymptotic states.
We conclude this first inspection of the state of the two-level system mentioning that, although for slow enough relative motion there is a preferred time interval in which the effect of velocity can be drew out from the net vacuum induced effect, as the velocity is increased and its effect is reflected in the mix asymptotic state, long time studies could supply evidence of the velocity induced phenomenons as well.

From the obtained reduced density matrix, we have estimated the decoherence timescale at which the coherences are strongly suppressed. We have also provided a low velocity expression for this timescale, which has been shown to scale as $u^2$ in accordance to previous results, and we have further analyzed its dependence on a variety of parameters involved in the dynamics. Through analytic considerations, we have shown how both, the net effect of the composite environment on the particle and the velocity dependent effect are strongly dependent on the material parameters and the system level spacing, allowing to amplify or weaken the magnitude by a sensible choice. 

Finally, as for the dependence upon the polarization direction, we have found  results for the decoherence time in agreement  with those existing in the Literature for Quantum Friction, showing a qualitative inverse proportionality among them and the frictional force.  This means that a link between the decoherence time and the quantum frictional force can be established since non-contact friction seems to enhance the decoherence of the moving atom. This suggests that measuring decoherence times could be used to indirectly demonstrate the existence of quantum friction. Furthermore, we have made an analysis of the different compositions of the material and effective two-level system that can be useful in experimental setups with the purpose of enhancing the effect of the finite velocity on the internal degree of the particle.  In particular, we found the rate between the decoherence timescale at finite velocity and that at null velocity to be bigger by a factor $10^2$ when comparing an NV center moving over a n-Si surface with a Rb atom moving over a gold surface.
Luckily our work can spark renewed optimism in the design of new experimental setups for the detection of  non-contact friction with the hope that this non-equilibrium phenomenon can be sighted in measurable reality soon.
\section*{Acknowledgements}

This work was supported by ANPCyT, CONICET, and Universidad de Buenos Aires; Argentina. 

\appendix
\section{Master equation coefficients }\label{appendixa}
   The polarization vector of the particle can be split in module and direction as $\mathbf{d}=d (n_x\, \check{x}+n_y\, \check{y}+n_z\, \check{z})$, with $\sum n_i^2 = 1$. Then, coefficients $D(v,t)$, $f(v,t)$ and $\zeta(v,t)$ appearing in the master equation are given, in polar coordinates for the parallel wave vector $\mathbf{k}$, by expressions of the form
    
\begin{align}\nonumber
     N(v,t)=\frac{d^2}{4 \pi^2 \hbar }\int_0^t dt'\; \int_0^{2\pi}d\theta_k \; \int_0^{\infty} \;k^2\, dk\,d\omega\;e^{-2ak} \\[.75em]\nonumber
  \times   [\left(\text{n}_x\cos(\theta_k) + \text{n}_y\sin(\theta_k)\right)^2 + \text{n}_z^2]
     \frac{\Gamma \omega_p^2 \; \omega}{\left(\omega^2-\omega_s^2\right)^2+\Gamma^2\omega^2}\\[.75em]\nonumber
      \text{trig}(\Delta(t-t'))\text{trig}\left((\omega-\mathfrak{i}k v \cos(\theta_k))(t-t')\right)
\end{align}

where we referring to any of the coefficient functions as $N(v,t)$ and  either function sin($x$) or cos($x$) as ${\rm trig}$.

Expressing the trigonometric function $\text{trig}\left((\omega-\mathfrak{i}k v \cos(\theta_k))(t-t')\right)$ in terms of exponential functions, the integral over $k$ can be performed directly

\begin{equation*}
    \int_0^{\infty } k^2 e^{-2ak} e^ {\pm i k v \cos (\theta ) (t-t')} \, dk=\frac{2}{(2 a \mp i v \cos (\theta ) (t-t'))^3},
\end{equation*}

as well as the non zero integrals over $\theta_k$,

\begin{align*}
    \int_0^{2 \pi }  d\theta\frac{\cos ^2(\theta )}{(2 a \pm i v \cos (\theta ) (t-t'))^3} &=\frac{2 \pi  \left(2 a^2-v^2 (t-t')^2\right)}{\left(4 a^2+v^2 (t-t')^2\right)^{5/2}}\\[.75em]
    \int_0^{2 \pi }  d\theta\frac{\sin ^2(\theta )}{(2 a \pm i v \cos (\theta ) (t-t'))^3}  &=\frac{\pi }{\left(4 a^2+v^2 (t-t')^2\right)^{3/2}}\\[.75em]
    \int_0^{2 \pi }  d\theta\frac{1}{(2 a \pm i v \cos (\theta ) (t-t'))^3}  &=\frac{\pi  \left(8 a^2-v^2 (t-t')^2\right)}{\left(4 a^2+v^2 (t-t')^2\right)^{5/2}}
\end{align*}

Then, if we define dimensionless parameters 
\begin{equation}
    u=\frac{v}{\omega_{\text{s}}\times a}\quad ;\quad \Tilde{\Delta}=\frac{\Delta}{\omega_{\text{s}}}\quad ;\quad \Tilde{\Gamma}=\frac{\Gamma}{\omega_{\text{s}}},
\end{equation}
the dimensional coefficient $r_0 = (d^2\; \omega_{\text{p}}^2)/(\hbar\;\omega_{\text{s}}^2\;a^3)$, and change to dimensionless variables 

\begin{equation}
    \frac{\omega}{\omega_{\text{s}}} \rightarrow \omega\quad ;\quad \frac{\Gamma}{\omega_{\text{s}}}\rightarrow t,
\end{equation}
the functions can be written as
\begin{widetext}
\begin{align}
    D(v,t)&=  \frac{r_0}{2\pi}\int_0^t dt'\;\int_0^\infty d\omega \frac{\tilde{\Gamma}\; \omega}{\left(\omega^2-1\right)^2+\tilde{\Gamma}^2\omega^2} \cos(\tilde{\Delta}t')\cos(\omega t') \mathbf{P}(ut') \label{D_raw}\\
    f(v,t)&= \frac{r_0}{2\pi}\int_0^t dt'\;\int_0^\infty d\omega \frac{\tilde{\Gamma}\; \omega}{\left(\omega^2-1\right)^2+\tilde{\Gamma}^2\omega^2} \sin(\tilde{\Delta}t')\cos(\omega t') \mathbf{P}(ut')\label{f_raw}\\
    \zeta(v,t)&=\frac{r_0}{2\pi}\int_0^t dt'\;\int_0^\infty d\omega \frac{\tilde{\Gamma}\; \omega}{\left(\omega^2-1\right)^2+\tilde{\Gamma}^2\omega^2} \sin(\tilde{\Delta}t')\sin(\omega t') \mathbf{P}(ut'),\label{Z_raw}
\end{align}
\end{widetext}
where $\mathbf{P}(ut')$ is an algebraic function given by

\begin{eqnarray}
    \mathbf{P}(ut') &=& 2\text{n}_x^2\frac{2 -u^2 t'^2}{\left(4 +u^2 t'^2\right)^{5/2}} \nonumber \\ &+&  \frac{\text{n}_y^2 }{\left(4 +u^2 t'^2\right)^{3/2}}  + \text{n}_z^2\frac{  \left(8-u^2 t'^2\right)}{\left(4 +u^2 t'^2\right)^{5/2}}.
\end{eqnarray}

In the following, we will only consider $D(v,t)$, since the other coefficients can be treated in a very similar manner. The expression to be integrated over $\omega$,

\begin{equation}
    \frac{1}{2}\int_0^\infty\;d\omega \frac{\tilde{\Gamma}\; \omega}{\left(\omega^2-1\right)^2+\tilde{\Gamma}^2\omega^2} \left(e^{\mathfrak{i}\omega t'} + e^{-\mathfrak{i}\omega t'}\right),
    \label{w_integal}
\end{equation}
 is holomorphic everywhere but on the poles given by the roots of the denominator $\{\tilde{\omega}_{\text{r}},-\tilde{\omega}_{\text{r}},c.c.\}$, with
\begin{equation}
    \tilde{\omega}_{\text{r}}= \frac{1}{\sqrt{2}}\sqrt{2-\tilde{\Gamma} + \mathfrak{i}\sqrt{4-\tilde{\Gamma}}}.
    \label{raiz}
\end{equation}

For $\tilde{\Gamma}$ values satisfying $\tilde{\Gamma}<2$ these are complex poles with both real and imaginary non-vanishing parts, while for $\tilde{\Gamma}>2$ $\tilde{\omega}_{\text{r}}$ is purely imaginary.
Integral (\ref{w_integal}) can be expressed in terms of exponential integral functions as 
\begin{align}
    &\frac{1}{4\sqrt{4-\tilde{\Gamma}^2}}\times\\[0.75em]\nonumber
    &\left(\pi \, e^{\mathfrak{i}\omega_{\text{r}}t}-\mathfrak{i}e^{\mathfrak{i}\omega_{\text{r}}t}\text{E}_1(\mathfrak{i}\omega_{\text{r}}t)-\mathfrak{i}e^{-\mathfrak{i}\omega_{\text{r}}t}\text{E}_1(-\mathfrak{i}\omega_{\text{r}}t) +\; \text{c.c.}\right),
\end{align}
where the terms containing the functions $\text{E}_1$ are negligible when compared to the exponential term, as long as $\tilde{\Gamma} << 1$ but cease to be negligible for $\tilde{\Gamma} \sim 1$.

Firstly, we will focus in the approximated result obtained for small $\tilde{\Gamma}$ values (i.e. disregarding all therms containing $\text{E}_1$ functions) to explain how we have considered the correction introduced by them at the end of this appendix.
At this point, we are only left with the time integral to be addressed. In order to do so, we shall extend $t'$ to the complex plane and modify the integration path conveniently. By observing that for non-relativistic velocities of the particle, oscillations in
\begin{align}
     &D(v,t)=\frac{r_0}{4\pi}\frac{\pi}{\sqrt{4 - \tilde{\Gamma^2}}}  \\[.75em]\nonumber
     &\hspace{1.25cm}\int_0^t dt' \;\left(e^{i\, (\tilde{\omega}_{\text{r}} + \tilde{\Delta})t'} + e^{i\, (\tilde{\omega}_{\text{r}} - \tilde{\Delta})t'}+ \text{c.c.}\right)\mathbf{P}(ut')
\end{align}
occur in a much faster timescale than $\mathbf{P}(ut')$ variation, we can adopt steepest descent method to replace the integrals over the real axis intervals by integrals along constant phase path for each exponential.
\begin{figure}[h]
\centering
\includegraphics[width=.95\columnwidth]{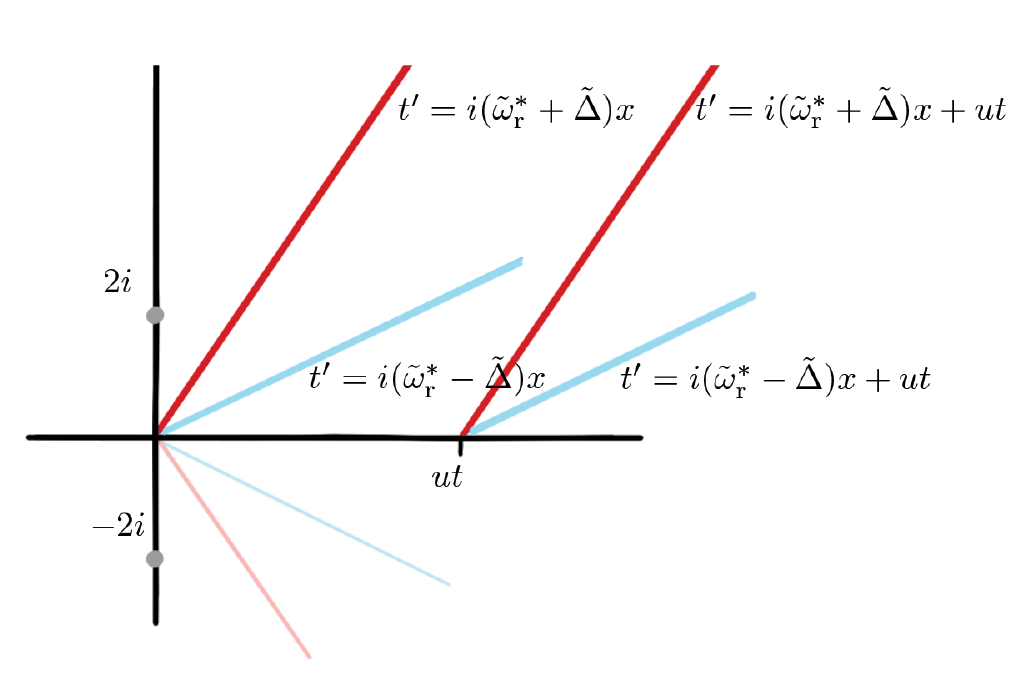}
\caption{Constant phase contours for integral over the variable t' which was extended to complex plane.\label{contorno_t}}
\end{figure}
Over these paths, the integral is dominated by the contributions at those points where the exponent is a local maximum, allowing for an expansion of the algebraic function in the parameter of the curve. Following this method we get, up to second order in the adimensional velocity $u$, the approximated expression
\begin{align}
     &D(v,t)\sim \frac{r_0}{8}\frac{1}{\sqrt{4 - \tilde{\Gamma^2}}}  \label{orientacion} \\[.75em]\nonumber
     &\left[-\frac{d^{(i)}}{8}\Im\frac{1}{\tilde{\omega}_{\text{r}} + \tilde{\Delta}}\right.- \frac{3}{32}d^{(a)} \, u^2 \Im\frac{1}{(\tilde{\omega}_{\text{r}} + \tilde{\Delta})^3} \\ \nonumber
     &+ \mathbf{P}(ut') \Im\frac{e^{i\, (\tilde{\omega}_{\text{r}} + \tilde{\Delta})t'}}{\tilde{\omega}_{\text{r}} + \tilde{\Delta}} + 12\; \mathbf{R}(ut)\,u^2\;\Im\frac{e^{i\, (\tilde{\omega}_{\text{r}} + \tilde{\Delta})t'}}{(\tilde{\omega}_{\text{r}} + \tilde{\Delta})^3}\\ \nonumber
     &\left.- 3\;\mathbf{Q}(ut)t \,u^2\;\Re\frac{e^{i\, (\tilde{\omega}_{\text{r}} + \tilde{\Delta})t'}}{(\tilde{\omega}_{\text{r}} + \tilde{\Delta})^2} \right] + (\tilde{\Delta}\leftrightarrow-\tilde{\Delta}),
\end{align}
where $d^{(i)} = 1 + \text{n}_z^2$, $d^{(a)}=3 \text{n}_x^2 +\text{n}_y^2+4\text{n}_z^2$. The additional algebraic functions \textbf{Q} and \textbf{R} appearing in this expression for $D(v,t)$ are given by
\begin{align*}
    \mathbf{Q}&=2\frac{6 -u^2 t^2}{\left(4 +u^2 t^2\right)^{7/2}}\; \text{n}_x^2 +\frac{\text{n}_y^2}{\left(4 +u^2 t^2\right)^{5/2}}+\frac{16 -u^2 t^2}{\left(4 +u^2 t^2\right)^{7/2}}\;\text{n}_z^2\\
    \mathbf{R}&=2\frac{(6 -u^2 t^2)^2-30}{\left(4 +u^2 t^2\right)^{9/2}}\; \text{n}_x^2 +\frac{(1-u^2t^2)}{\left(4 +u^2 t^2\right)^{7/2}}\;\text{n}_y^2\\
    &\hspace{0cm}+\frac{16 -27u^2 t^2+u^4t^4}{\left(4 +u^2 t^2\right)^{9/2}}\;\text{n}_z^2,
\end{align*}
while approximate solutions can be found for $f(v,t)$ and $\zeta(v,t)$ following an analogous procedure.

In order to incorporate a correction that allows to investigate greater $\tilde{\Gamma}$ values, we expand $\mathbf{P}(ut)$ up to second order in $u$ in the integrand, so that
\begin{equation}
    \mathbf{P}(ut')\sim \frac{d^{(i)}}{8} - \frac{3}{64}d^{(a)} \, u^2 t'^2.
\end{equation}
Allowing to write the remaining part of the integral as
\begin{align}\nonumber
     &\Delta D(v,t)=\frac{r_0}{16\pi}\frac{1}{\sqrt{4 - \tilde{\Gamma^2}}} \left[\frac{d^{(i)}}{8} + \frac{3}{64}d^{(a)} \, u^2 \partial_{\tilde{\Delta}}^2\right] \\[.75em]
     &\hspace{1.5cm}\int_0^t dt'\;\left\lbrace(-\mathfrak{i})\left(e^{\mathfrak{i}(\omega_{\text{r}} + \tilde{\Delta})t}\;\text{E}_1(\mathfrak{i}\omega_{\text{r}}t)+\right.\right.\\[.75em]\nonumber
     &\left.\left. \hspace{1cm} e^{-\mathfrak{i}(\omega_{\text{r}}+\tilde{\Delta})t}\;\text{E}_1(-\mathfrak{i}\omega_{\text{r}}t)  \right)+ \text{c.c.}\right\rbrace + (\tilde{\Delta} \leftrightarrow -\tilde{\Delta}).
\end{align}

This integral can be formally solved to

\begin{align}\nonumber
     &\Delta D(v,t)=\frac{r_0}{16\pi}\frac{1}{\sqrt{4 - \tilde{\Gamma^2}}} \left[\frac{d^{(i)}}{8} + \frac{3}{64}d^{(a)} \, u^2 \partial_{\tilde{\Delta}}^2\right] \\[.75em]
     &\left\lbrace\frac{-2\pi \mathfrak{i}}{\tilde{\omega}_{\text{r}} + \tilde{\Delta}} + \left[ (\text{E}_1(-\mathfrak{i}\tilde{\Delta}t)-\text{E}_1(\mathfrak{i}\tilde{\Delta}t)) \frac{1}{\tilde{\omega}_{\text{r}} + \tilde{\Delta}}     
     \right.\right.\\[.75em]\nonumber
     & -\frac{e^{\mathfrak{i}(\omega_{\text{r}} + \tilde{\Delta})t}}{\tilde{\omega}_{\text{r}} + \tilde{\Delta}}\;\text{E}_1(\mathfrak{i}\omega_{\text{r}}t)+ \frac{e^{-\mathfrak{i}(\omega_{\text{r}} + \tilde{\Delta})t}}{\tilde{\omega}_{\text{r}} + \tilde{\Delta}}\;\text{E}_1(-\mathfrak{i}\omega_{\text{r}}t) \\[.75em]\nonumber
     &\hspace{4.5cm}\left.\left. + (\tilde{\Delta} \leftrightarrow -\tilde{\Delta})\right] + \text{c.c.}\right\rbrace
\end{align}

\begin{figure}[h!]
\centering
\includegraphics[width=.95\columnwidth, trim= 10 0 10 0]{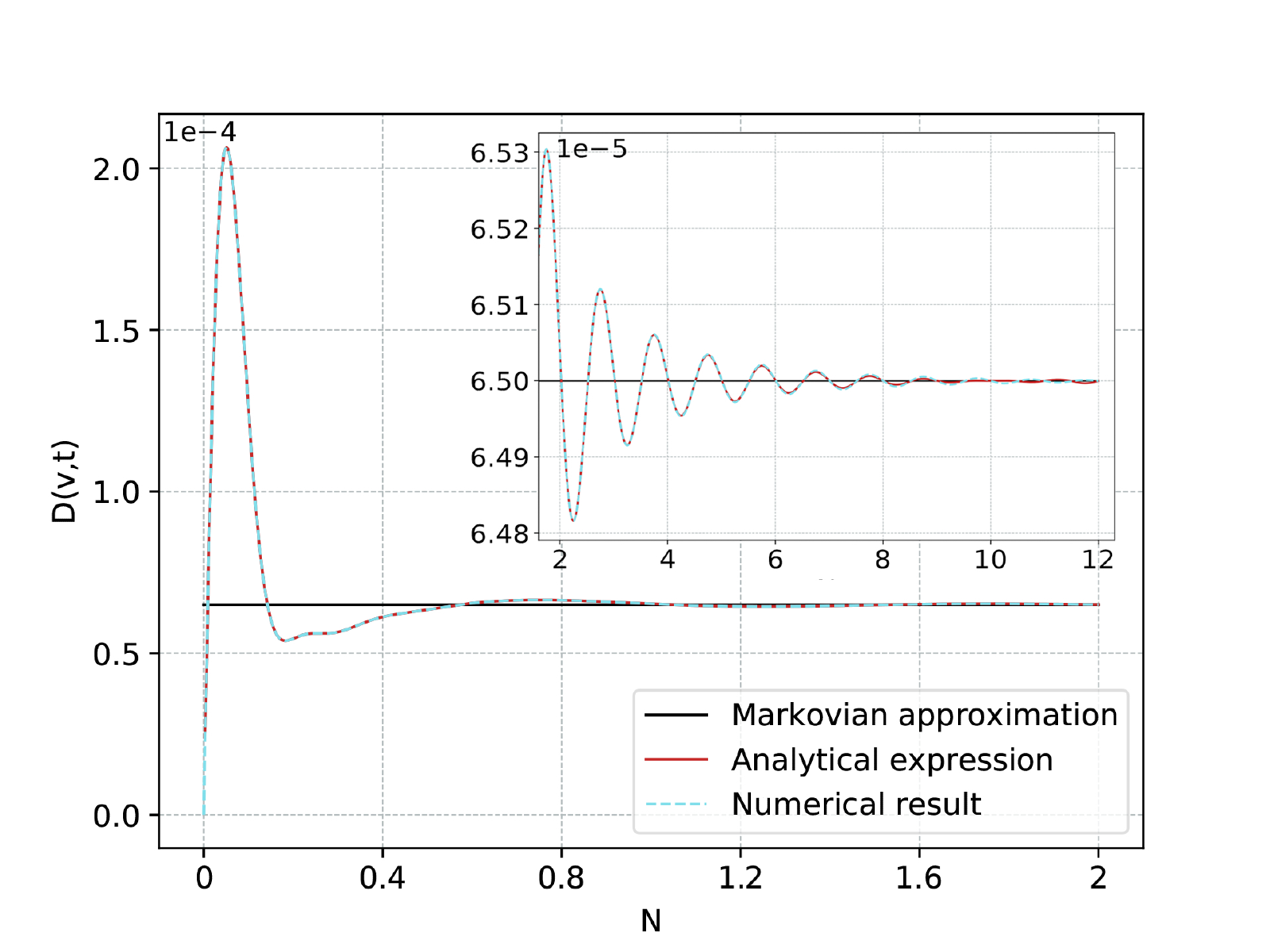}
\caption{Coefficient D(v,t) evolution in natural cycles $N = \frac{t}{2\pi/\tilde{\Delta}}$, comparing the analytical expression, the numerical result and the value obtained when performing Markov approximation. Parameters values are $a=5$nm, $\tilde{\Gamma}=1$, $\tilde{\Delta}=0.2$, $u=0.003$ and $\mathbf{d}=d(1,0,0)$.\label{D_ingles}}
\end{figure}
These expressions has been tested against the numerical results. An easy comparison is possible to be obtained from Fig. \ref{D_ingles} where the evolution of $D(v,t)$ is plotted for $\tilde{\Gamma} = 1$ The dashed line, representing the numerical result is not distinguishable from the red solid line representing the analytical expression. Both the numerical and the analytical results tend to its markovian approximation value for a few number of cycles for these parameter values, but a difference on the value accumulated when computing $\int D(v,t)$ and $D^{Markov}\times t$ can be suspected.\\

\section{Dependence on level spacing}\label{appendixb}
\begin{figure}[h!]
\centering
\includegraphics[width=.9\columnwidth, trim= 20 0 20 0]{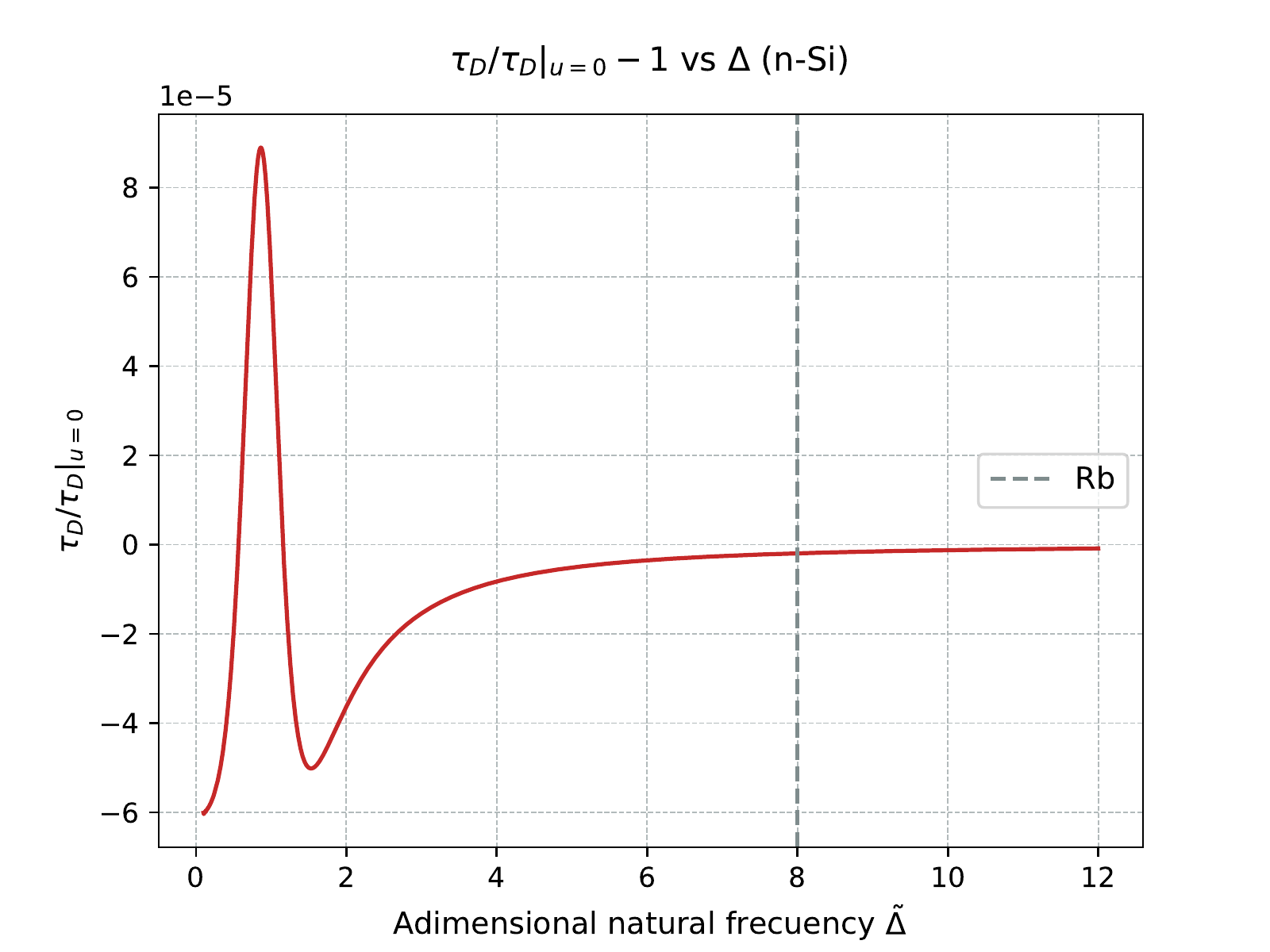}
\includegraphics[width=.9\columnwidth, trim= 20 0 20 0]{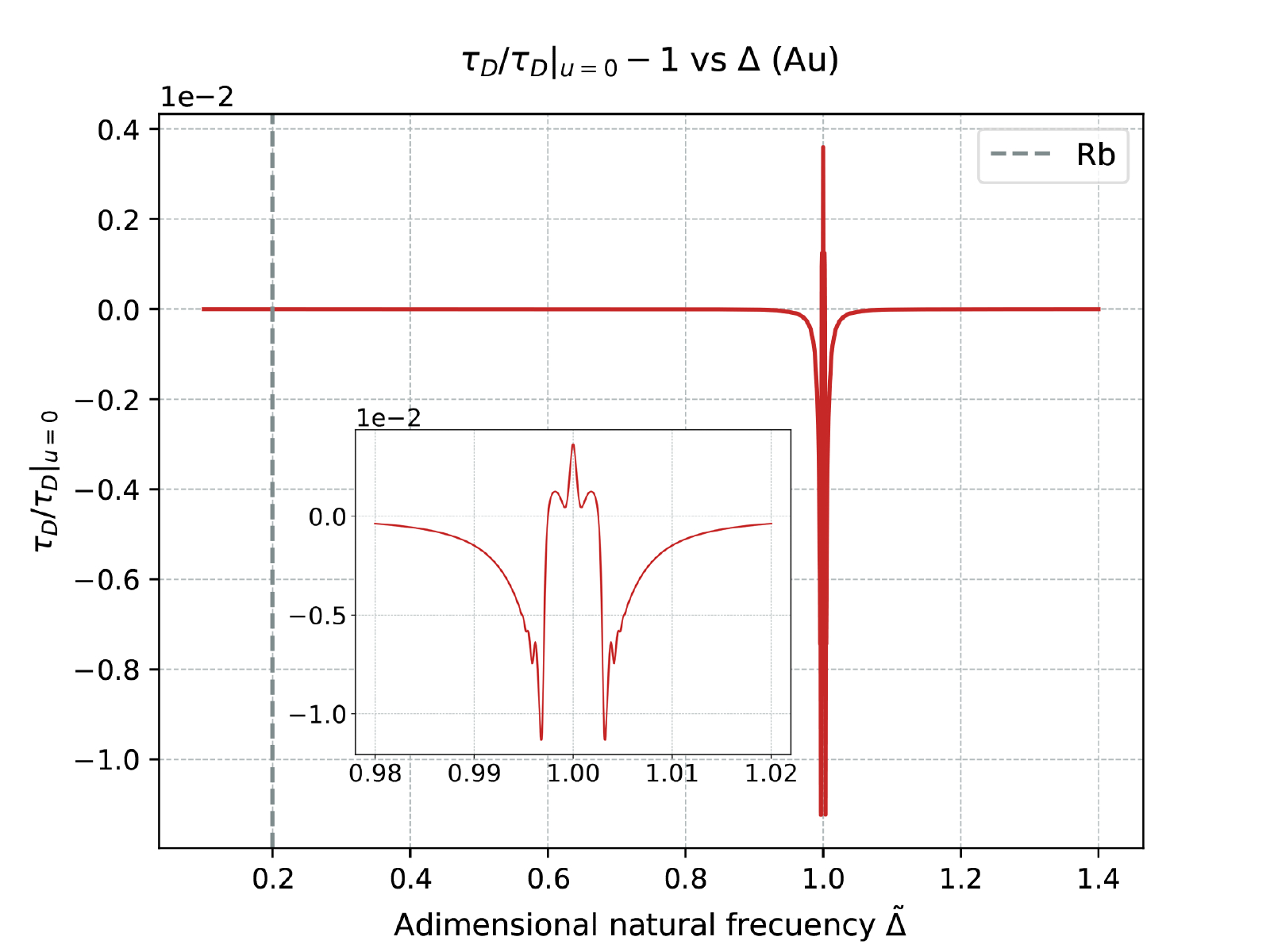}
\caption{Decoherence time as a function of the dimensionless level spacing $\tilde{\Delta}$ of the system, normalized with the null velocity value, considering an n-Si (up) an a gold (down) dielectric. Parameters values are $\tilde{\Gamma}=1$, $u=0.003$ for n-Si and $\tilde{\Gamma}=3\times 10^{-3}$, $u=1.5\times 10^{-4}$ and $\mathbf{d}=d(1,0,0)$.\label{tdec_vs_Delta}}
\end{figure}
As suggested in section \ref{destruction_seccion}, the strength of the velocity dependent effect can be studied from the relation between the decoherence time defined by Eq. (\ref{tdec_tlargo}) for finite and zero velocities.
Exploring the dependence of this relation with the level spacing of the atom, it can be seen in Fig. \ref{tdec_vs_Delta} that an appropriated choice of the particle to be considered can enhance (or diminished) the effect of velocity considerably. Rigorously speaking, the dependence is not on the level spacing alone but on the relation between the Drude-Lorentz parameters of the material and the level spacing. It can be seen that, moving apart from the prohibited near-resonance range, the difference between the vacuum and velocity effects is maximum for certain values of natural level spacing $\tilde{\Delta}$, approaching zero sufficiently far from resonance.
In order to maximize the effect to be studied, we will consider the system to be an NV center of natural frequency $\tilde{\Delta}=0.2$ moving over an n doped silicon surface or an NV center of natural frequency $\tilde{\Delta}=0.9$ moving over a gold surface unless explicitly indicated.

\bibliography{0casimir.bib,0cuantica.bib,0teoria.bib}

\begin{thebibliography}{46}
\expandafter\ifx\csname natexlab\endcsname\relax\def\natexlab#1{#1}\fi
\expandafter\ifx\csname bibnamefont\endcsname\relax
  \def\bibnamefont#1{#1}\fi
\expandafter\ifx\csname bibfnamefont\endcsname\relax
  \def\bibfnamefont#1{#1}\fi
\expandafter\ifx\csname citenamefont\endcsname\relax
  \def\citenamefont#1{#1}\fi
\expandafter\ifx\csname url\endcsname\relax
  \def\url#1{\texttt{#1}}\fi
\expandafter\ifx\csname urlprefix\endcsname\relax\def\urlprefix{URL }\fi
\providecommand{\bibinfo}[2]{#2}
\providecommand{\eprint}[2][]{\url{#2}}

\bibitem[{\citenamefont{Casimir}(1948)}]{Casimir}
\bibinfo{author}{\bibfnamefont{H.~B.} \bibnamefont{Casimir}}, in
  \emph{\bibinfo{booktitle}{Proceedings of the KNAW}} (\bibinfo{year}{1948}),
  vol.~\bibinfo{volume}{51}, pp. \bibinfo{pages}{793--795}.

\bibitem[{\citenamefont{Milonni}(2013)}]{book_milonni}
\bibinfo{author}{\bibfnamefont{P.~W.} \bibnamefont{Milonni}},
  \emph{\bibinfo{title}{The quantum vacuum: an introduction to quantum
  electrodynamics}} (\bibinfo{publisher}{Academic press},
  \bibinfo{year}{2013}).

\bibitem[{\citenamefont{Lamoreaux}(1997)}]{Lamoreaux1997}
\bibinfo{author}{\bibfnamefont{S.~K.} \bibnamefont{Lamoreaux}},
  \bibinfo{journal}{Phys. Rev. Lett.} \textbf{\bibinfo{volume}{78}},
  \bibinfo{pages}{5} (\bibinfo{year}{1997}).

\bibitem[{\citenamefont{Bordag et~al.}(2009)\citenamefont{Bordag,
  Klimchitskaya, Mohideen, and Mostepanenko}}]{bordag1}
\bibinfo{author}{\bibfnamefont{M.}~\bibnamefont{Bordag}},
  \bibinfo{author}{\bibfnamefont{G.}~\bibnamefont{Klimchitskaya}},
  \bibinfo{author}{\bibfnamefont{U.}~\bibnamefont{Mohideen}}, \bibnamefont{and}
  \bibinfo{author}{\bibfnamefont{V.}~\bibnamefont{Mostepanenko}},
  \emph{\bibinfo{title}{{Advances in the Casimir effect}}}, vol.
  \bibinfo{volume}{145} (\bibinfo{publisher}{Oxford University Press},
  \bibinfo{year}{2009}).

\bibitem[{\citenamefont{Bordag et~al.}(2001)\citenamefont{Bordag, Mohideen, and
  Mostepanenko}}]{bordag2}
\bibinfo{author}{\bibfnamefont{M.}~\bibnamefont{Bordag}},
  \bibinfo{author}{\bibfnamefont{U.}~\bibnamefont{Mohideen}}, \bibnamefont{and}
  \bibinfo{author}{\bibfnamefont{V.}~\bibnamefont{Mostepanenko}},
  \bibinfo{journal}{Physics Reports} \textbf{\bibinfo{volume}{353}},
  \bibinfo{pages}{1 } (\bibinfo{year}{2001}), ISSN \bibinfo{issn}{0370-1573}.

\bibitem[{\citenamefont{Milton}(2001)}]{book_milton}
\bibinfo{author}{\bibfnamefont{K.~A.} \bibnamefont{Milton}},
  \emph{\bibinfo{title}{The Casimir effect: physical manifestations of
  zero-point energy}} (\bibinfo{publisher}{World Scientific},
  \bibinfo{year}{2001}).

\bibitem[{\citenamefont{Milton}(2004)}]{milton2004casimir}
\bibinfo{author}{\bibfnamefont{K.~A.} \bibnamefont{Milton}},
  \bibinfo{journal}{J. Phys. A} \textbf{\bibinfo{volume}{37}},
  \bibinfo{pages}{R209} (\bibinfo{year}{2004}).

\bibitem[{\citenamefont{Reynaud et~al.}(2001)\citenamefont{Reynaud, Lambrecht,
  Genet, and Jaekel}}]{reynaud2001quantum}
\bibinfo{author}{\bibfnamefont{S.}~\bibnamefont{Reynaud}},
  \bibinfo{author}{\bibfnamefont{A.}~\bibnamefont{Lambrecht}},
  \bibinfo{author}{\bibfnamefont{C.}~\bibnamefont{Genet}}, \bibnamefont{and}
  \bibinfo{author}{\bibfnamefont{M.-T.} \bibnamefont{Jaekel}},
  \bibinfo{journal}{Comptes Rendus Acad. Sci.-Series IV-Phys.}
  \textbf{\bibinfo{volume}{2}}, \bibinfo{pages}{1287} (\bibinfo{year}{2001}).

\bibitem[{\citenamefont{Dalvit et~al.}(2011)\citenamefont{Dalvit, Neto, and
  Mazzitelli}}]{review_friction}
\bibinfo{author}{\bibfnamefont{D.~A.} \bibnamefont{Dalvit}},
  \bibinfo{author}{\bibfnamefont{P.~A.~M.} \bibnamefont{Neto}},
  \bibnamefont{and} \bibinfo{author}{\bibfnamefont{F.~D.}
  \bibnamefont{Mazzitelli}}, \bibinfo{journal}{Lec. Notes Phys.}
  \textbf{\bibinfo{volume}{834}}, \bibinfo{pages}{419} (\bibinfo{year}{2011}).

\bibitem[{\citenamefont{Nation et~al.}(2012)\citenamefont{Nation, Johansson,
  Blencowe, and Nori}}]{nation_colloquium}
\bibinfo{author}{\bibfnamefont{P.~D.} \bibnamefont{Nation}},
  \bibinfo{author}{\bibfnamefont{J.~R.} \bibnamefont{Johansson}},
  \bibinfo{author}{\bibfnamefont{M.~P.} \bibnamefont{Blencowe}},
  \bibnamefont{and} \bibinfo{author}{\bibfnamefont{F.}~\bibnamefont{Nori}},
  \bibinfo{journal}{Rev. Mod. Phys.} \textbf{\bibinfo{volume}{84}},
  \bibinfo{pages}{1} (\bibinfo{year}{2012}).

\bibitem[{\citenamefont{Wilson et~al.}(2011)\citenamefont{Wilson, Johansson,
  Pourkabirian, Simoen, Johansson, Duty, Nori, and
  Delsing}}]{dyncasexp_supercond}
\bibinfo{author}{\bibfnamefont{C.}~\bibnamefont{Wilson}},
  \bibinfo{author}{\bibfnamefont{G.}~\bibnamefont{Johansson}},
  \bibinfo{author}{\bibfnamefont{A.}~\bibnamefont{Pourkabirian}},
  \bibinfo{author}{\bibfnamefont{M.}~\bibnamefont{Simoen}},
  \bibinfo{author}{\bibfnamefont{J.}~\bibnamefont{Johansson}},
  \bibinfo{author}{\bibfnamefont{T.}~\bibnamefont{Duty}},
  \bibinfo{author}{\bibfnamefont{F.}~\bibnamefont{Nori}}, \bibnamefont{and}
  \bibinfo{author}{\bibfnamefont{P.}~\bibnamefont{Delsing}},
  \bibinfo{journal}{Nature} \textbf{\bibinfo{volume}{479}},
  \bibinfo{pages}{376} (\bibinfo{year}{2011}).

\bibitem[{\citenamefont{L{\"a}hteenm{\"a}ki
  et~al.}(2013)\citenamefont{L{\"a}hteenm{\"a}ki, Paraoanu, Hassel, and
  Hakonen}}]{dyncasexp_squid}
\bibinfo{author}{\bibfnamefont{P.}~\bibnamefont{L{\"a}hteenm{\"a}ki}},
  \bibinfo{author}{\bibfnamefont{G.}~\bibnamefont{Paraoanu}},
  \bibinfo{author}{\bibfnamefont{J.}~\bibnamefont{Hassel}}, \bibnamefont{and}
  \bibinfo{author}{\bibfnamefont{P.~J.} \bibnamefont{Hakonen}},
  \bibinfo{journal}{Proc. Natl. Acad. Sci.} \textbf{\bibinfo{volume}{110}},
  \bibinfo{pages}{4234} (\bibinfo{year}{2013}).

\bibitem[{\citenamefont{Pendry}(1997)}]{pendry97}
\bibinfo{author}{\bibfnamefont{J.}~\bibnamefont{Pendry}}, \bibinfo{journal}{J.
  Phys.} \textbf{\bibinfo{volume}{9}}, \bibinfo{pages}{10301}
  (\bibinfo{year}{1997}).

\bibitem[{\citenamefont{Pendry}(2010{\natexlab{a}})}]{pendry_debate}
\bibinfo{author}{\bibfnamefont{J.}~\bibnamefont{Pendry}}, \bibinfo{journal}{New
  J. Phys.} \textbf{\bibinfo{volume}{12}}, \bibinfo{pages}{033028}
  (\bibinfo{year}{2010}{\natexlab{a}}).

\bibitem[{\citenamefont{Pendry}(2010{\natexlab{b}})}]{Pendry_reply}
\bibinfo{author}{\bibfnamefont{J.~B.} \bibnamefont{Pendry}},
  \bibinfo{journal}{New Journal of Physics} \textbf{\bibinfo{volume}{12}},
  \bibinfo{pages}{068002} (\bibinfo{year}{2010}{\natexlab{b}}).

\bibitem[{\citenamefont{Leonhardt}(2010)}]{Leonhardt_2010}
\bibinfo{author}{\bibfnamefont{U.}~\bibnamefont{Leonhardt}},
  \bibinfo{journal}{New Journal of Physics} \textbf{\bibinfo{volume}{12}},
  \bibinfo{pages}{068001} (\bibinfo{year}{2010}).

\bibitem[{\citenamefont{Philbin and Leonhardt}(2009)}]{Leonhardt_nofriction}
\bibinfo{author}{\bibfnamefont{T.}~\bibnamefont{Philbin}} \bibnamefont{and}
  \bibinfo{author}{\bibfnamefont{U.}~\bibnamefont{Leonhardt}},
  \bibinfo{journal}{New Journal of Physics} \textbf{\bibinfo{volume}{11}}
  (\bibinfo{year}{2009}), ISSN \bibinfo{issn}{1367-2630}.

\bibitem[{\citenamefont{Volokitin and Persson}(2007)}]{volokitin_persson}
\bibinfo{author}{\bibfnamefont{A.}~\bibnamefont{Volokitin}} \bibnamefont{and}
  \bibinfo{author}{\bibfnamefont{B.}~\bibnamefont{Persson}},
  \bibinfo{journal}{Rev. Mod. Phys.} \textbf{\bibinfo{volume}{79}},
  \bibinfo{pages}{1291} (\bibinfo{year}{2007}).

\bibitem[{\citenamefont{Klatt et~al.}(2017)\citenamefont{Klatt, Far\'{\i}as,
  Dalvit, and Buhmann}}]{dalvit_arbitrarily}
\bibinfo{author}{\bibfnamefont{J.}~\bibnamefont{Klatt}},
  \bibinfo{author}{\bibfnamefont{M.~B.} \bibnamefont{Far\'{\i}as}},
  \bibinfo{author}{\bibfnamefont{D.~A.~R.} \bibnamefont{Dalvit}},
  \bibnamefont{and} \bibinfo{author}{\bibfnamefont{S.~Y.}
  \bibnamefont{Buhmann}}, \bibinfo{journal}{Phys. Rev. A}
  \textbf{\bibinfo{volume}{95}}, \bibinfo{pages}{052510}
  (\bibinfo{year}{2017}).

\bibitem[{\citenamefont{Barton}(2010)}]{barton_atom_halfspace}
\bibinfo{author}{\bibfnamefont{G.}~\bibnamefont{Barton}}, \bibinfo{journal}{New
  J. Phys.} \textbf{\bibinfo{volume}{12}}, \bibinfo{pages}{113045}
  (\bibinfo{year}{2010}).

\bibitem[{\citenamefont{Intravaia et~al.}(2015)\citenamefont{Intravaia,
  Mkrtchian, Buhmann, Stefan, Dalvit, and Henkel}}]{intravaia_acceleration}
\bibinfo{author}{\bibfnamefont{F.}~\bibnamefont{Intravaia}},
  \bibinfo{author}{\bibfnamefont{V.~E.} \bibnamefont{Mkrtchian}},
  \bibinfo{author}{\bibfnamefont{S.~Y.} \bibnamefont{Buhmann}},
  \bibinfo{author}{\bibnamefont{Stefan}}, \bibinfo{author}{\bibfnamefont{D.~A.}
  \bibnamefont{Dalvit}}, \bibnamefont{and}
  \bibinfo{author}{\bibfnamefont{C.}~\bibnamefont{Henkel}},
  \bibinfo{journal}{J. Phys. Condens. Matter} \textbf{\bibinfo{volume}{27}},
  \bibinfo{pages}{214020} (\bibinfo{year}{2015}).

\bibitem[{\citenamefont{Scheel and Buhmann}(2009)}]{Scheel}
\bibinfo{author}{\bibfnamefont{S.}~\bibnamefont{Scheel}} \bibnamefont{and}
  \bibinfo{author}{\bibfnamefont{S.~Y.} \bibnamefont{Buhmann}},
  \bibinfo{journal}{Phys. Rev. A} \textbf{\bibinfo{volume}{80}},
  \bibinfo{pages}{042902} (\bibinfo{year}{2009}),
  \urlprefix\url{https://link.aps.org/doi/10.1103/PhysRevA.80.042902}.

\bibitem[{\citenamefont{Intravaia et~al.}(2014)\citenamefont{Intravaia,
  Behunin, and Dalvit}}]{dalvit_fluctuation}
\bibinfo{author}{\bibfnamefont{F.}~\bibnamefont{Intravaia}},
  \bibinfo{author}{\bibfnamefont{R.}~\bibnamefont{Behunin}}, \bibnamefont{and}
  \bibinfo{author}{\bibfnamefont{D.}~\bibnamefont{Dalvit}},
  \bibinfo{journal}{Phys. Rev. A} \textbf{\bibinfo{volume}{89}},
  \bibinfo{pages}{050101(R)} (\bibinfo{year}{2014}).

\bibitem[{\citenamefont{Reiche et~al.}(2020)\citenamefont{Reiche, Intravaia,
  Hsiang, Busch, and Hu}}]{hu_thermodynamic}
\bibinfo{author}{\bibfnamefont{D.}~\bibnamefont{Reiche}},
  \bibinfo{author}{\bibfnamefont{F.}~\bibnamefont{Intravaia}},
  \bibinfo{author}{\bibfnamefont{J.-T.} \bibnamefont{Hsiang}},
  \bibinfo{author}{\bibfnamefont{K.}~\bibnamefont{Busch}}, \bibnamefont{and}
  \bibinfo{author}{\bibfnamefont{B.~L.} \bibnamefont{Hu}},
  \bibinfo{journal}{Phys. Rev. A} \textbf{\bibinfo{volume}{102}},
  \bibinfo{pages}{050203(R)} (\bibinfo{year}{2020}).

\bibitem[{\citenamefont{Farias et~al.}(2018)\citenamefont{Farias, Kort-Kamp,
  and Dalvit}}]{farias2018quantum}
\bibinfo{author}{\bibfnamefont{M.~B.} \bibnamefont{Farias}},
  \bibinfo{author}{\bibfnamefont{W.~J.~M.} \bibnamefont{Kort-Kamp}},
  \bibnamefont{and} \bibinfo{author}{\bibfnamefont{D.~A.~R.}
  \bibnamefont{Dalvit}}, \bibinfo{journal}{Phys. Rev. B}
  \textbf{\bibinfo{volume}{97}}, \bibinfo{pages}{161407(R)}
  (\bibinfo{year}{2018}).

\bibitem[{\citenamefont{Marino et~al.}(2017)\citenamefont{Marino, Recati, and
  Carusotto}}]{carusotto2017friction}
\bibinfo{author}{\bibfnamefont{J.}~\bibnamefont{Marino}},
  \bibinfo{author}{\bibfnamefont{A.}~\bibnamefont{Recati}}, \bibnamefont{and}
  \bibinfo{author}{\bibfnamefont{I.}~\bibnamefont{Carusotto}},
  \bibinfo{journal}{Phys. Rev. Lett.} \textbf{\bibinfo{volume}{118}},
  \bibinfo{pages}{045301} (\bibinfo{year}{2017}).

\bibitem[{\citenamefont{Intravaia et~al.}(2019)\citenamefont{Intravaia,
  Oelschl\"ager, Reiche, Dalvit, and Busch}}]{intravaia_rolling}
\bibinfo{author}{\bibfnamefont{F.}~\bibnamefont{Intravaia}},
  \bibinfo{author}{\bibfnamefont{M.}~\bibnamefont{Oelschl\"ager}},
  \bibinfo{author}{\bibfnamefont{D.}~\bibnamefont{Reiche}},
  \bibinfo{author}{\bibfnamefont{D.~A.~R.} \bibnamefont{Dalvit}},
  \bibnamefont{and} \bibinfo{author}{\bibfnamefont{K.}~\bibnamefont{Busch}},
  \bibinfo{journal}{Phys. Rev. Lett.} \textbf{\bibinfo{volume}{123}},
  \bibinfo{pages}{120401} (\bibinfo{year}{2019}).

\bibitem[{\citenamefont{Farias et~al.}(2017)\citenamefont{Farias, Fosco,
  Lombardo, and Mazzitelli}}]{farias_graphene}
\bibinfo{author}{\bibfnamefont{M.~B.} \bibnamefont{Farias}},
  \bibinfo{author}{\bibfnamefont{C.~D.} \bibnamefont{Fosco}},
  \bibinfo{author}{\bibfnamefont{F.~C.} \bibnamefont{Lombardo}},
  \bibnamefont{and} \bibinfo{author}{\bibfnamefont{F.~D.}
  \bibnamefont{Mazzitelli}}, \bibinfo{journal}{Phys. Rev. D}
  \textbf{\bibinfo{volume}{95}}, \bibinfo{pages}{065012}
  (\bibinfo{year}{2017}).

\bibitem[{\citenamefont{Far{\'\i}as et~al.}(2015)\citenamefont{Far{\'\i}as,
  Fosco, Lombardo, Mazzitelli, and L{\'o}pez}}]{farias_friction}
\bibinfo{author}{\bibfnamefont{M.~B.} \bibnamefont{Far{\'\i}as}},
  \bibinfo{author}{\bibfnamefont{C.~D.} \bibnamefont{Fosco}},
  \bibinfo{author}{\bibfnamefont{F.~C.} \bibnamefont{Lombardo}},
  \bibinfo{author}{\bibfnamefont{F.~D.} \bibnamefont{Mazzitelli}},
  \bibnamefont{and} \bibinfo{author}{\bibfnamefont{A.~E.~R.}
  \bibnamefont{L{\'o}pez}}, \bibinfo{journal}{Phys. Rev. D}
  \textbf{\bibinfo{volume}{91}}, \bibinfo{pages}{105020}
  (\bibinfo{year}{2015}).

\bibitem[{\citenamefont{Viotti et~al.}(2019)\citenamefont{Viotti,
  Bel\'en~Far\'{\i}as, Villar, and Lombardo}}]{viotti_thermal}
\bibinfo{author}{\bibfnamefont{L.}~\bibnamefont{Viotti}},
  \bibinfo{author}{\bibfnamefont{M.}~\bibnamefont{Bel\'en~Far\'{\i}as}},
  \bibinfo{author}{\bibfnamefont{P.~I.} \bibnamefont{Villar}},
  \bibnamefont{and} \bibinfo{author}{\bibfnamefont{F.~C.}
  \bibnamefont{Lombardo}}, \bibinfo{journal}{Phys. Rev. D}
  \textbf{\bibinfo{volume}{99}}, \bibinfo{pages}{105005}
  (\bibinfo{year}{2019}).

\bibitem[{\citenamefont{Volokitin and Persson}(2011)}]{volokitin2011quantum}
\bibinfo{author}{\bibfnamefont{A.}~\bibnamefont{Volokitin}} \bibnamefont{and}
  \bibinfo{author}{\bibfnamefont{B.}~\bibnamefont{Persson}},
  \bibinfo{journal}{Phys. Rev. Lett.} \textbf{\bibinfo{volume}{106}},
  \bibinfo{pages}{094502} (\bibinfo{year}{2011}).

\bibitem[{\citenamefont{Volokitin and Persson}(2016)}]{volokitin_cherenkov}
\bibinfo{author}{\bibfnamefont{A.~I.} \bibnamefont{Volokitin}}
  \bibnamefont{and} \bibinfo{author}{\bibfnamefont{B.~N.~Y.}
  \bibnamefont{Persson}}, \bibinfo{journal}{JETP Letters}
  \textbf{\bibinfo{volume}{103}} (\bibinfo{year}{2016}).

\bibitem[{\citenamefont{Klatt et~al.}(2016)\citenamefont{Klatt, Bennett, and
  Buhmann}}]{buhmann_spectroscopic}
\bibinfo{author}{\bibfnamefont{J.}~\bibnamefont{Klatt}},
  \bibinfo{author}{\bibfnamefont{R.}~\bibnamefont{Bennett}}, \bibnamefont{and}
  \bibinfo{author}{\bibfnamefont{S.~Y.} \bibnamefont{Buhmann}},
  \bibinfo{journal}{Phys. Rev. A} \textbf{\bibinfo{volume}{94}},
  \bibinfo{pages}{063803} (\bibinfo{year}{2016}).

\bibitem[{\citenamefont{Far{\'\i}as and Lombardo}(2016)}]{farias_decoherence}
\bibinfo{author}{\bibfnamefont{M.~B.} \bibnamefont{Far{\'\i}as}}
  \bibnamefont{and} \bibinfo{author}{\bibfnamefont{F.~C.}
  \bibnamefont{Lombardo}}, \bibinfo{journal}{Phys. Rev. D}
  \textbf{\bibinfo{volume}{93}}, \bibinfo{pages}{065035}
  (\bibinfo{year}{2016}).

\bibitem[{\citenamefont{Farías et~al.}(2020)\citenamefont{Farías, Lombardo,
  Soba, Villar, and Decca}}]{farias_nature}
\bibinfo{author}{\bibfnamefont{M.~B.} \bibnamefont{Farías}},
  \bibinfo{author}{\bibfnamefont{F.~C.} \bibnamefont{Lombardo}},
  \bibinfo{author}{\bibfnamefont{A.}~\bibnamefont{Soba}},
  \bibinfo{author}{\bibfnamefont{P.~I.} \bibnamefont{Villar}},
  \bibnamefont{and} \bibinfo{author}{\bibfnamefont{R.~S.} \bibnamefont{Decca}},
  \bibinfo{journal}{npj Quantum Information} \textbf{\bibinfo{volume}{6}},
  \bibinfo{pages}{25} (\bibinfo{year}{2020}).

\bibitem[{\citenamefont{Barton}(1997)}]{barton78}
\bibinfo{author}{\bibfnamefont{G.}~\bibnamefont{Barton}},
  \bibinfo{journal}{Proceedings of the Royal Society of London. Series A:
  Mathematical, Physical and Engineering Sciences}
  \textbf{\bibinfo{volume}{453}}, \bibinfo{pages}{2461} (\bibinfo{year}{1997}).

\bibitem[{\citenamefont{Barton}(1979)}]{barton79}
\bibinfo{author}{\bibfnamefont{G.}~\bibnamefont{Barton}},
  \bibinfo{journal}{Reports on Progress in Physics}
  \textbf{\bibinfo{volume}{42}}, \bibinfo{pages}{963} (\bibinfo{year}{1979}).

\bibitem[{\citenamefont{Breuer and Petruccione}(2002)}]{petruccione}
\bibinfo{author}{\bibfnamefont{H.-P.} \bibnamefont{Breuer}} \bibnamefont{and}
  \bibinfo{author}{\bibfnamefont{F.}~\bibnamefont{Petruccione}},
  \emph{\bibinfo{title}{The theory of open quantum systems}}
  (\bibinfo{publisher}{Oxford University Press on Demand},
  \bibinfo{year}{2002}).

\bibitem[{\citenamefont{Paz and Zurek}(2001)}]{paz_mastereq}
\bibinfo{author}{\bibfnamefont{J.~P.} \bibnamefont{Paz}} \bibnamefont{and}
  \bibinfo{author}{\bibfnamefont{W.~H.} \bibnamefont{Zurek}}, in
  \emph{\bibinfo{booktitle}{Coherent atomic matter waves}}, edited by
  \bibinfo{editor}{\bibfnamefont{R.}~\bibnamefont{Kaiser}},
  \bibinfo{editor}{\bibfnamefont{C.}~\bibnamefont{Westbrook}},
  \bibnamefont{and} \bibinfo{editor}{\bibfnamefont{F.}~\bibnamefont{David}}
  (\bibinfo{publisher}{Springer Berlin Heidelberg}, \bibinfo{address}{Berlin,
  Heidelberg}, \bibinfo{year}{2001}), pp. \bibinfo{pages}{533--614}, ISBN
  \bibinfo{isbn}{978-3-540-45338-3}.

\bibitem[{\citenamefont{Leggett et~al.}(1987)\citenamefont{Leggett,
  Chakravarty, Dorsey, Fisher, Garg, and Zwerger}}]{Leggett}
\bibinfo{author}{\bibfnamefont{A.~J.} \bibnamefont{Leggett}},
  \bibinfo{author}{\bibfnamefont{S.}~\bibnamefont{Chakravarty}},
  \bibinfo{author}{\bibfnamefont{A.~T.} \bibnamefont{Dorsey}},
  \bibinfo{author}{\bibfnamefont{M.~P.~A.} \bibnamefont{Fisher}},
  \bibinfo{author}{\bibfnamefont{A.}~\bibnamefont{Garg}}, \bibnamefont{and}
  \bibinfo{author}{\bibfnamefont{W.}~\bibnamefont{Zwerger}},
  \bibinfo{journal}{Rev. Mod. Phys.} \textbf{\bibinfo{volume}{59}},
  \bibinfo{pages}{1} (\bibinfo{year}{1987}).

\bibitem[{\citenamefont{Dattagupta and Puri}(2004)}]{book_referee2}
\bibinfo{author}{\bibfnamefont{S.}~\bibnamefont{Dattagupta}} \bibnamefont{and}
  \bibinfo{author}{\bibfnamefont{S.}~\bibnamefont{Puri}},
  \emph{\bibinfo{title}{Dissipative Phenomena in Condensed Matter. Some
  Applications}} (\bibinfo{publisher}{Springer, Berlin, Heidelberg},
  \bibinfo{year}{2004}).

\bibitem[{\citenamefont{Fleming et~al.}(2010)\citenamefont{Fleming, Cummings,
  Anastopoulos, and Hu}}]{hu_rwa}
\bibinfo{author}{\bibfnamefont{C.}~\bibnamefont{Fleming}},
  \bibinfo{author}{\bibfnamefont{N.~I.} \bibnamefont{Cummings}},
  \bibinfo{author}{\bibfnamefont{C.}~\bibnamefont{Anastopoulos}},
  \bibnamefont{and} \bibinfo{author}{\bibfnamefont{B.~L.} \bibnamefont{Hu}},
  \bibinfo{journal}{Journal of Physics A: Mathematical and Theoretical}
  \textbf{\bibinfo{volume}{43}}, \bibinfo{pages}{405304}
  (\bibinfo{year}{2010}).

\bibitem[{\citenamefont{Haikka and Maniscalco}(2010)}]{maniscalco3}
\bibinfo{author}{\bibfnamefont{P.}~\bibnamefont{Haikka}} \bibnamefont{and}
  \bibinfo{author}{\bibfnamefont{S.}~\bibnamefont{Maniscalco}},
  \bibinfo{journal}{Phys. Rev. A} \textbf{\bibinfo{volume}{81}},
  \bibinfo{pages}{052103} (\bibinfo{year}{2010}).

\bibitem[{\citenamefont{Intravaia et~al.}(2016)\citenamefont{Intravaia,
  Behunin, Henkel, Busch, and Dalvit}}]{dalvit_nonmarkovianity}
\bibinfo{author}{\bibfnamefont{F.}~\bibnamefont{Intravaia}},
  \bibinfo{author}{\bibfnamefont{R.~O.} \bibnamefont{Behunin}},
  \bibinfo{author}{\bibfnamefont{C.}~\bibnamefont{Henkel}},
  \bibinfo{author}{\bibfnamefont{K.}~\bibnamefont{Busch}}, \bibnamefont{and}
  \bibinfo{author}{\bibfnamefont{D.~A.~R.} \bibnamefont{Dalvit}},
  \bibinfo{journal}{Phys. Rev. A} \textbf{\bibinfo{volume}{94}},
  \bibinfo{pages}{042114} (\bibinfo{year}{2016}).

\bibitem[{\citenamefont{Svidzinsky}(2019)}]{plasmones}
\bibinfo{author}{\bibfnamefont{A.~A.} \bibnamefont{Svidzinsky}},
  \bibinfo{journal}{Phys. Rev. Research} \textbf{\bibinfo{volume}{1}},
  \bibinfo{pages}{033027} (\bibinfo{year}{2019}).

\bibitem[{\citenamefont{Maghrebi et~al.}(2013)\citenamefont{Maghrebi,
  Golestanian, and Kardar}}]{cherenkov}
\bibinfo{author}{\bibfnamefont{M.~F.} \bibnamefont{Maghrebi}},
  \bibinfo{author}{\bibfnamefont{R.}~\bibnamefont{Golestanian}},
  \bibnamefont{and} \bibinfo{author}{\bibfnamefont{M.}~\bibnamefont{Kardar}},
  \bibinfo{journal}{Phys. Rev. A} \textbf{\bibinfo{volume}{88}},
  \bibinfo{pages}{042509} (\bibinfo{year}{2013}).

\end{thebibliography}




\end{document}